%% file: main.tex
\documentclass[conference]{IEEEtran}
\IEEEoverridecommandlockouts

\usepackage{cite}
\usepackage{graphicx} % Required for inserting images
\usepackage{xcolor}
\usepackage{comment}
\usepackage{multicol}
\usepackage{amsfonts}
\usepackage{subcaption}
\usepackage{theorem}
\usepackage{multicol}
\usepackage{listings}
\usepackage{url}

\newtheorem{theorem}{Observation}[section]

\begin{document}

%\author{Sajal Dash, Isaac Lyngaas, Junqi Yin, Xiao Wang, 

%Romain Egele, Guojing Cong, Feiyi Wang, Prasanna Balaprakash}
%\begin{comment}
\author{\IEEEauthorblockN{Sajal Dash}
\IEEEauthorblockA{%\textit{National Center for Computational Sciences} \\
\textit{\small Oak Ridge National Laboratory}\\
\small dashs@ornl.gov}
\and
\IEEEauthorblockN{Isaac R Lyngaas}
\IEEEauthorblockA{%\textit{National Center for Computational Sciences} \\
\textit{\small Oak Ridge National Laboratory}\\
\small lyngaasir@ornl.gov}
\and
\IEEEauthorblockN{Junqi Yin}
\IEEEauthorblockA{%\textit{National Center for Computational Sciences} \\
\textit{\small Oak Ridge National Laboratory}\\
\small yinj@ornl.gov}
\and
\IEEEauthorblockN{Xiao Wang}
\IEEEauthorblockA{%\textit{Computational Sciences \& Engineering Division} \\
\textit{Oak Ridge National Laboratory}\\
\small wangx2@ornl.gov}
\and
\IEEEauthorblockN{Romain Egele}
\IEEEauthorblockA{%\textit{National Center for Computational Sciences} \\
\textit{\small Universit\'{e} Paris-Saclay}\\
\small romainegele@gmail.com}
\and
\IEEEauthorblockN{Guojing Cong}
\IEEEauthorblockA{%\textit{National Center for Computational Sciences} \\
\textit{\small Oak Ridge National Laboratory}\\
\small congg@ornl.gov}
\and
\IEEEauthorblockN{Feiyi Wang}
\IEEEauthorblockA{%\textit{National Center for Computational Sciences} \\
\textit{\small Oak Ridge National Laboratory}\\
\small fwang2@ornl.gov }
\and
\IEEEauthorblockN{Prasanna Balaprakash}
\IEEEauthorblockA{%\textit{National Center for Computational Sciences} \\
\textit{Oak Ridge National Laboratory}\\
\small pbalapra@ornl.gov }}
%\end{comment}

\title{Optimizing Distributed Training on Frontier for Large Language Models}

\maketitle

\input{abstract}

\input{introduction}

\input{background}

%\subsection{Analysis of Frontier System and Its Compute Capability}
\input{porting}

\input{empirical-analysis}

\input{deephyper}

\input{recipe}
\input{conclusion}

\section*{Acknowledgements}
This research was sponsored by and used resources of the Oak Ridge Leadership Computing Facility (OLCF), which is a DOE Office of Science User Facility at the Oak Ridge National Laboratory supported by the U.S. Department of Energy under Contract No. DE-AC05-00OR22725.

%\clearpage

%\begin{thebibliography}{00}

\bibliographystyle{IEEEtran}
\bibliography{main}
%\end{thebibliography}

%\clearpage
%\onecolumn
%\appendix
%\input{counting-flops}
\end{document}

%% file: abstract.tex
\begin{abstract}
Large language models (LLMs) have demonstrated remarkable success as foundational models, benefiting various downstream applications through fine-tuning. Recent studies on loss scaling have demonstrated the superior performance of larger LLMs compared to their smaller counterparts. Nevertheless, training LLMs with billions of parameters poses significant challenges and requires considerable computational resources. For example, training a one trillion parameter GPT-style model on 20 trillion tokens requires a staggering 120 million exaflops of computation. This research explores efficient distributed training strategies to extract this computation from Frontier, the world's first exascale supercomputer dedicated to open science. We enable and investigate various model and data parallel training techniques, such as tensor parallelism, pipeline parallelism, and sharded data parallelism, to facilitate training a trillion-parameter model on Frontier. We empirically assess these techniques and their associated parameters to determine their impact on memory footprint, communication latency, and GPU's computational efficiency. We analyze the complex interplay among these techniques and find a strategy to combine them to achieve high throughput through hyperparameter tuning. We have identified efficient strategies for training large LLMs of varying sizes through empirical analysis and hyperparameter tuning. For 22 Billion, 175 Billion, and 1 Trillion parameters, we achieved GPU throughputs of $38.38\%$, $36.14\%$, and $31.96\%$, respectively. For the training of the 175 Billion parameter model and the 1 Trillion parameter model, we achieved $100\%$ weak scaling efficiency on 1024 and 3072 MI250X GPUs, respectively. We also achieved strong scaling efficiencies of $89\%$ and $87\%$ for these two models. 
\end{abstract}

%% file: introduction.tex
\section{Introduction}
Large language models (LLMs) leverage an attention mechanism to learn language structure and can generate natural language responses to many prompts. Once trained on a large corpus of text, these models can be fine-tuned to perform many downstream tasks; thus, LLMs are very successful as foundational models. Recent studies demonstrated that LLM models with a large number of parameters outperform LLM models with a smaller number of parameters~\cite{kaplan2020scaling}. Large LLMs such as GPT3-175B~\cite{brown2020language}, BLOOM-176B~\cite{workshop2022bloom}, OPT-175B~\cite{zhang2022opt}, and Turing NLG-530B~\cite{smith2022using} have shown remarkable success as foundational models and outperform their smaller counterparts in many NLP tasks. Some studies also reported the loss scaling law, which states that an LLM model can keep learning from data up to 20x-200x of its parameter count~\cite{kaplan2020scaling, hoffmann2022training, yin2023forge}. Training large models using large data requires a tremendous amount of computing resources. Cost and energy-efficient utilization of these resources is always challenging.

These models' success stories demonstrate that open-sourced large models can serve as state-of-the-art foundation models. With the advent of RedPajama datasets with one Trillion and 30 Trillion tokens~\cite{together2023redpajama}, and the Dolma dataset with three Trillion tokens~\cite{DolmaDataset, DolmaToolkit}, a model of size 1 Trillion parameter must be within the horizon. A rough estimate~\cite{yin2023evaluation} tells us that training a Trillion parameter model on 1-30 Trillion tokens will require $6\times 10^{12} \times [1-30]\times 10^{12} = 6-180$ Million exa-flops (floating point operations).

\begin{comment}
Large AI models can process and interpret large datasets, enabling them to tackle complex problems that were previously intractable. The learning capabilities have significantly increased in various applications such as natural language processing, image recognition, and predictive analytics. In the emerging landscape of artificial intelligence, large models with billions of parameters mark a transformative era. This evolution marks a pivotal shift in AI and HPC, underscoring the importance of interdependence between advanced computational architectures and algorithmic strategies. 
\end{comment}

This paper details our experience training such large LLM models with billions to trillion parameters on Oak Ridge National Laboratory's (ORNL) Frontier supercomputer, one of the world's most advanced HPC systems. Central to our narrative is the acknowledgment of HPC systems as more than mere facilitators of computing resources. The Frontier supercomputer, powered by advanced AMD GPUs, represents a paradigm shift in computational capabilities. However, training AI models at the trillion-parameter scale introduces unique challenges. These include balancing the extreme computational demands with memory constraints and optimizing inter-node communication to mitigate performance bottlenecks. Our research provides a detailed analysis of these challenges and the strategies employed to overcome them, offering insights into the intricacies of large-scale model training in an HPC environment.

Large language models often hit GPU memory walls, and training a trillion parameter model requires ~24 Terabytes of memory. So, to fit this model, we need to break it down into parts and distribute them across hundreds of GPUs. LLMs are transformer models whose shapes are determined linearly by the depth (number of layers) and quadratically by the width (hidden dimension). Various model parallelization approaches distribute the model along these two dimensions. We can also use data parallelism to speed up training by utilizing more GPUs for training on large datasets. 

Tensor parallelism proposed by Megatron-LM~\cite{narayanan2021efficient} partitions the layer weights (Tensors) row-wise, performs computation on the smaller matrices, and combines the activation results. This approach splits the model across the width dimension. Pipeline parallelism breaks the model across layer dimensions and places groups of layers on individual GPUs. The micro-batches are consumed in a pipelined fashion, and backward and forward propagation of these micro-bathes are overlapped so that the communication latency can be hidden~\cite{huang2019gpipe,narayanan2019pipedream,li2021chimera}. The most traditional way for data parallelism is to replicate the entire model across GPU groups and train these replicas in parallel while averaging the loss after every forward pass. A novel direction of data parallelism, namely sharded data parallelism, achieves data parallelism by sharding the model parameters across available memory, reducing the amount of resources required for the model.

Megatron-DeepSpeed~\cite{megatron-deepspeed} supports tensor, pipeline, data, and sharded data parallelism. Megatron-LM supports the first three. DeepSpeed ZeRO~\cite{rajbhandari2020zero} and Fully Sharded Data Parallelism (FSDP)~\cite{zhao2023pytorch} support sharded data parallelism. Since models with trillion parameters require many modes of parallelism, Megatron-DeepSpeed is the most complete framework in terms of the different modes it supports. We explore this framework to find an optimal strategy through an investigative understanding of the complex interplay between these modes. However, these frameworks are primarily developed to run on NVIDIA GPUs and have yet to be tested extensively on a large scale or run on AMD platforms. So, we performed a feasibility study of running these frameworks on Frontier, ported the framework to Frontier to identify the limiting issues and opportunities, and prepared a training workflow with an optimized AI software stack.

After porting the Megatron-DeepSpeed framework to Frontier, we focus on finding the best strategies to train large models on Frontier using a combination of these parallelizations. The next challenge for a particular model is what combinations of these modes we should select and to what extent. For example, using tensor parallelism for wide models is beneficial. Still, because this incurs frequent communication after every layer, expanding tensor parallelism beyond the GPUs within a single node is not advised. For pipeline parallelism, the pipeline bubble can become an issue, making the communication latency a bottleneck. Finding the right pipeline parallelism parameters is crucial for hiding communication latency. So, we explore these modes and their right configuration in isolation and in combination.

\begin{comment}

Open-source distributed model training frameworks such as DeepSpeed and Megatron have emerged as cornerstones in reducing the complexity of training large models. 

\end{comment}
In our work, we delve into the specifics of how these tools are optimized for Frontier's AMD GPU architecture. This involves an in-depth exploration of their adaptability in managing extensive computational loads and memory optimization techniques necessary for training trillion-parameter models. The trade-offs between memory, compute, and communication are critical in the HPC context. Our research examines the interplay of these elements in the Frontier environment. We investigate how adjustments in distributed model training frameworks can be finely tuned to leverage the full potential of AMD GPUs, focusing on achieving an optimal balance between these components to maximize training efficiency and model accuracy.

A pivotal aspect of our study is the exploration of pipeline parallelism, tensor parallelism, micro-batch size, and gradient accumulation steps. These elements are fundamental in distributed training, particularly at the scale of trillion-parameter models. Our research provides a detailed examination of how each component is optimized for Frontier's infrastructure, focusing on how these parallelism strategies can be effectively implemented in an HPC setting to enhance computational efficiency and reduce training times.

%The Zero Redundancy Optimizer (Zero) plays a significant role in our approach, especially in the context of memory optimization. We present a comprehensive analysis of how the Zero optimizer is fine-tuned for Frontier's architecture, enabling more efficient utilization of memory resources and facilitating the training of larger models than previously possible.

In this paper, our primary focus lies in improving the training performance of these large-scale LLMs, particularly emphasizing the computational aspects and efficiency of various training strategies. It is important to note that our objective was not to train these models to completion for the purpose of achieving the highest possible accuracy. Instead, our approach was centered around understanding and enhancing the performance characteristics of training processes on HPC  systems. This involved exploring the scalability, efficiency, and resource utilization of different training methodologies, especially in the context of the Frontier supercomputer's capabilities. We seek to gain insights into how different parallelization techniques, model configurations, and system architectures impact the training dynamics of LLMs with billions to trillion parameters. Through this exploration, we sought to contribute practical strategies, which could aid in the efficient training of large-scale models in future research and practical applications.

\begin{comment}
Finally, we explore the scalability of model parallelism across various node configurations within the Frontier supercomputer. This involves an in-depth investigation into the effects of scaling model training across different numbers of nodes, analyzing how this impacts communication overhead, computational balance, and overall training efficiency. This aspect is crucial for understanding the dynamics of large-scale distributed training in an HPC setting, providing valuable insights into the optimization of inter-node communication and computational resource allocation.
\end{comment}

\subsection{Paper Outline}
Section~\ref{sec:background} discusses various distribution strategies and cost evaluation of training large LLMs on Frontier. Section~\ref{sec:empirical} provides an empirical analysis of multiple distribution strategies and associated parameters. We identify some valuable observations for training a 22B model from our experiments. In Section~\ref{sec:deephyper}, we report hyperparameter tuning for training a 175B model to understand the combinations of these distribution strategies. Section~\ref{sec:training} combines the lessons from Sections 3 and 4 and performs further experiments to devise a training recipe for 175B and 1T models. In that section, we also report GPU throughput, three different-sized models, and strong and weak scaling performance.

\subsection{Contributions}
The contributions of the paper are:

\begin{enumerate}
\item Distributed training techniques on AMD Hardware with ROCM software platform: This work %significantly
contributes to enabling %and optimizing 
state-of-the-art distributed training algorithms and frameworks for large LLMs %, specifically 175 billion and 1 trillion parameter models, 
on AMD hardware using the ROCM software platform. This advancement serves as a blueprint for efficient training of LLMs on non-NVIDIA and non-CUDA platforms, such as the AMD-powered Frontier supercomputer. %Our work addresses the gap in the current state-of-the-art frameworks that primarily target NVIDIA GPUs and are operated on CUDA-supported platforms.

\item Development of an optimized distributed training strategy through hyperparameter search: The research presents strategies to effectively manage the GPU memory wall and communication latency in the training of LLMs with billions to trillions of parameters. By performing empirical analysis and  hyperparameter search we identified a strategy that combines model parallelism techniques, such as tensor parallelism and pipeline parallelism, along with data parallelism to efficiently train large models of size 175 billion and 1 trillion parameters on Frontier. This study demonstrates how to optimize memory usage across multiple GPUs and minimize the communication latency. %This approach ensures efficient utilization of available resources, resulting in high GPU throughput and scaling efficiency.
\end{enumerate}

\begin{comment}
\begin{enumerate}
    \item We report our effort to port the state-of-the-art distributed training frameworks to Frontier.
    \item We performed hyperparameter tuning on distribution parameters to identify optimal combinations of various parallelization strategies on an AMD platform to train a Trillion parameter model.
    \item An optimum strategy for training the Trillion Parameter model on Frontier, the first exascale machine.
    \item We provide the first-ever training performance of the 1T model in scale on Frontier supercomputer.
    
\end{enumerate}
\end{comment}

%% file: background.tex
\section{Distributed Training Techniques and Frameworks}\label{sec:background}
Distributed training of large language models has seen many innovations and advances in recent times. Much of the focus on large language models has been on their ability to create more accurate models using an ever-increasing number of parameters. Due to this focus, models have become too large to fit in a single GPU's memory, accelerating the research into model parallelism techniques. Tensor parallelism, pipeline parallelism, and sharded data parallelism are the most popular techniques for model parallelism.

\subsection{GPT-style Model Architecture and Model Sizes}
\begin{figure}[ht]
    \centering
    \includegraphics[width=0.5\textwidth]{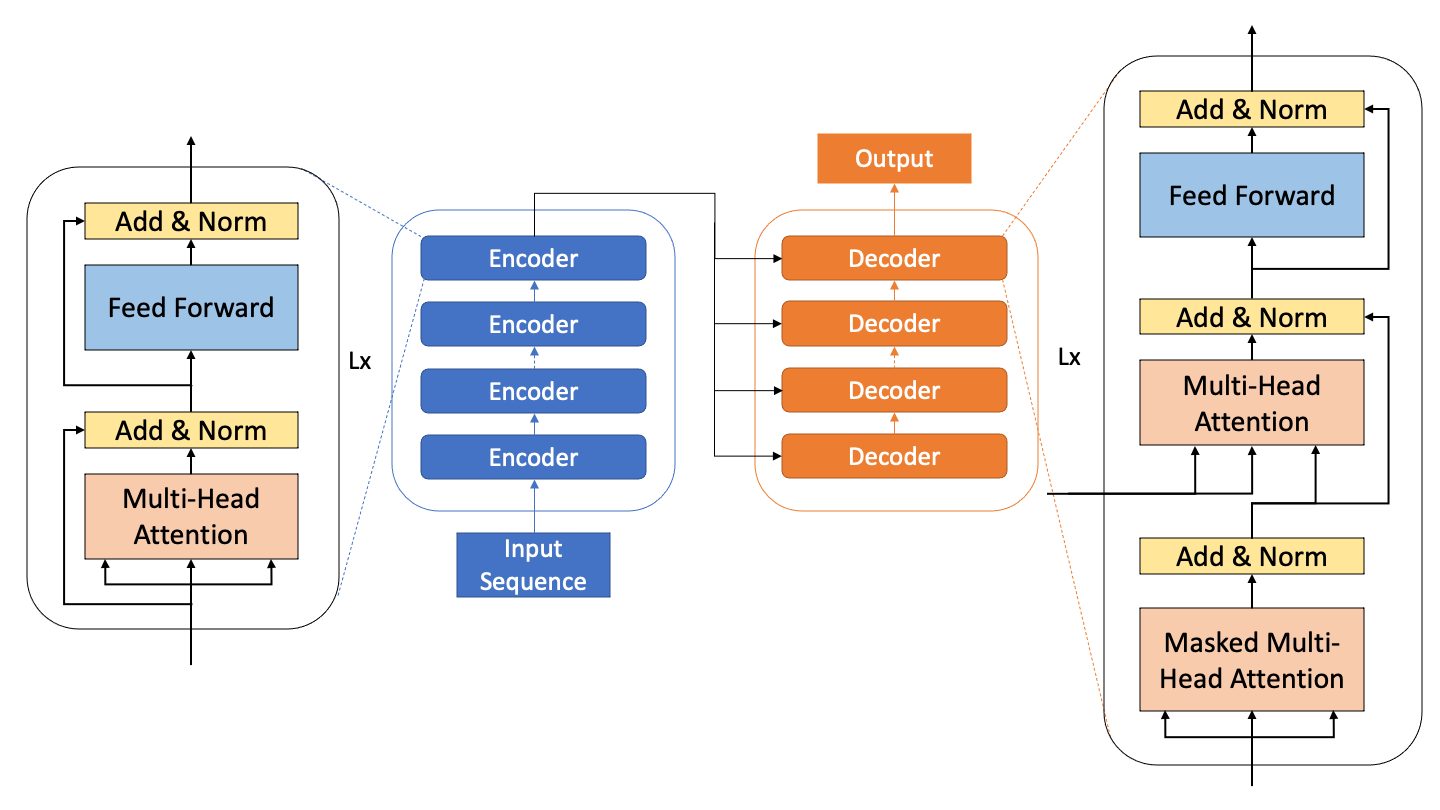}
    \caption{Transformer architecture.}
    \label{fig:transformer}
\end{figure}

Transformer models can consist of two distinct parts, a stack of encoder blocks and a stack of decoder blocks (Figure~\ref{fig:transformer})~\cite{NIPS2017_3f5ee243}. Encoder blocks help capture non-causal self-attention where each token in a sentence can attend to tokens from left and right. On the other hand, decoder blocks are useful for capturing causal self-attention, where a token can attend to only past tokens in the sequence. In a recent body of works, the encoder part has been co-opted for building BERT-like~\cite{devlin2018bert} models, which are useful for classification and regression types of work. On the other hand, the decoder block has been used for GPT-like~\cite{radford2018improving} models for generative tasks. 

\begin{figure}[ht]
    \centering
    \includegraphics[width=0.4\textwidth]{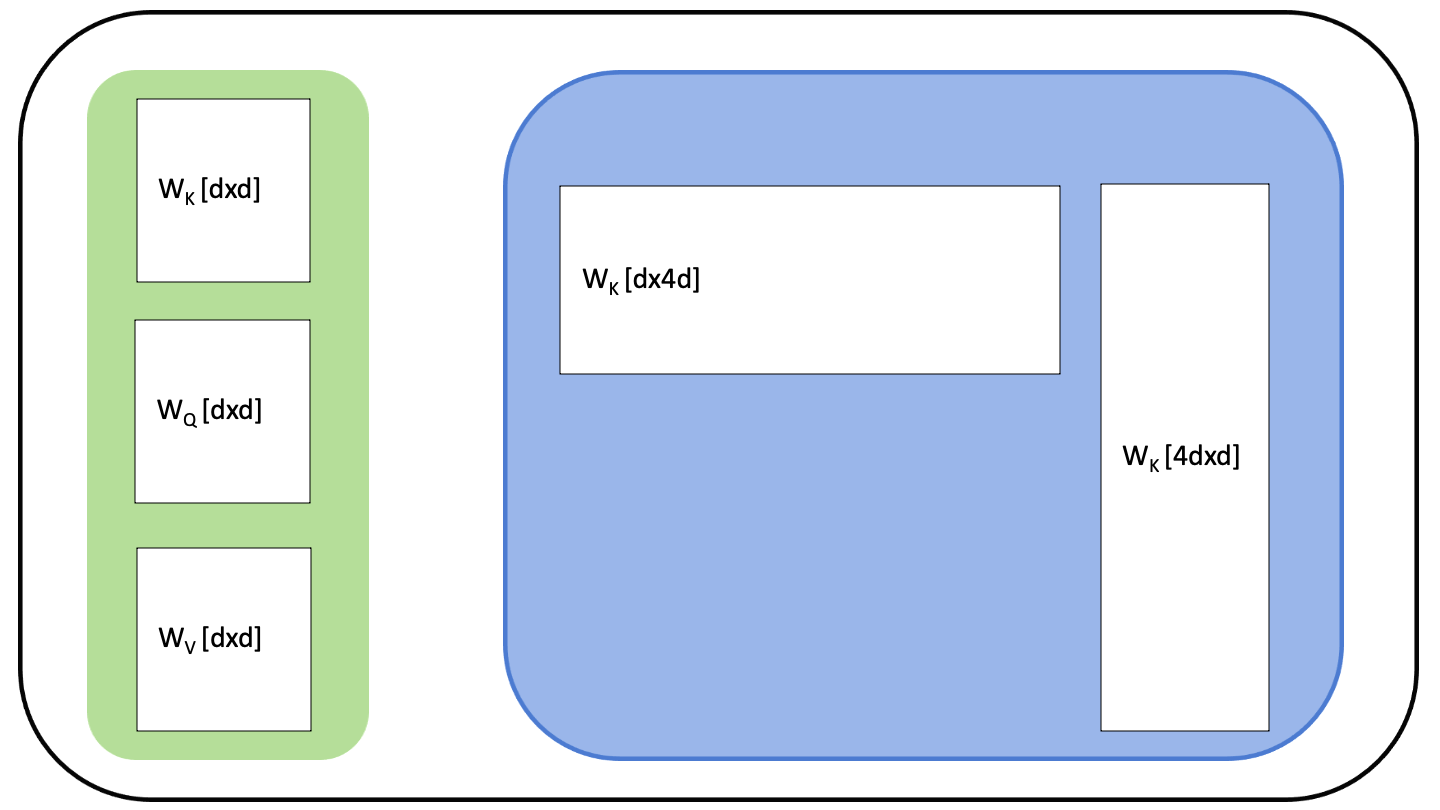}
    \caption{Model parameters of a GPT layer.}
    \label{fig:gpt-layer}
\end{figure}

The simplest GPT-like models consist of a stack of similar layers. Each layer has one attention block, followed by a Feed Forward Network (FFN)~\ref{fig:gpt-layer}. The attention block has three sets of parameters $W_K, W_Q, W_V \in \mathbb{R}^{d \times d}$, where $d$ is the hidden dimension of the models. The FFN block has two layers, with weights $W_1 \in \mathbb{R}^{d\times 4d}$ and  $W_2 \in \mathbb{R}^{4d\times d}$. So, a layer contributes to $11d^2$ parameters.

With the embedding layer at the beginning of the model, the number of parameters becomes roughly $12Ld^2$, where $L$ is the number of layers, and $d$ is the hidden dimension. With this formula, we can define three models with sizes 22B, 175B, and 1T in Table~\ref{tab:model-specs}.

\begin{table}[ht]
\centering

\begin{tabular}{cccc}
\hline
Model & \#Layers & Hidden size & \#Attention heads \\ \hline
1.4B  & 24 & 2114       & 24    \\ \hline
22B   & 48 & 6144        & 48      \\ \hline 
175B  & 96  & 12288      & 96        \\ \hline
1T    & 128 & 25600      & 128      \\ \hline
\end{tabular}
\caption{Architecture specification of GPT-style LLMs.}
\label{tab:model-specs}
\end{table}

Most memory requirements come from model weights, optimizer states, and gradients. However, the memory required for forward activation can also become significant depending on the batch size. In mixed precision training, we need 6 bytes for each model parameter, 4 to save the model in fp32, and 2 to use in computation in fp16. We need 4 bytes per parameter for Optimizer states to save the momentum in fp32 (Adam Optimizer). We need to save one fp32 gradient value for each parameter. So, in a mixed precision training with Adam optimizer, the minimum memory requirement is listed in Table~\ref{tab:mem-req}.

\begin{table}[h]
    \centering
    \begin{tabular}{c | c c c}
    \hline 
         &  \multicolumn{3}{c}{Memory Requirement}\\ \cline{2-4}
         Values & 22B Model & 175B Model & 1T Model \\ \hline \hline 
         Parameters (6x) & 132 GB & 1050 GB & 6 TB \\ \hline 
         Gradients (4x) & 88 GB & 700 GB & 4 TB \\ \hline 
         Optimizer States (4x) & 88 GB & 700 GB & 4 TB \\ \hline \hline 
         Total Memory (14x) & 308 GB & 2.45 TB & 14 TB \\ \hline 
    \end{tabular}
    \caption{Memory Requirement for Training a 22B, 175B, and 1T Models in mixed precision. We have not included the activation memory or the overhead memory for various distribution frameworks.}
    \label{tab:mem-req}
\end{table}

Each Frontier node has 8 MI250X GPUs\footnote{Each node has 4 MI250X cards, and each card has two Graphics Compute Dies (GCDs) or effective GPUs. Hence, going forward, we will use the term GPU to refer to the GCDs. Each GPU has a theoretical fp16 peak of 191.5 TFLOPS.}, each with 64 GB of HBM memory. So from Table~\ref{tab:model-specs}'s memory requirement, we can conclude that model parallelism is necessary to fit even one replica of the model. Model parallelism can happen in the hidden dimension via Tensor and Sharded data parallelism or in the layer dimension via Pipeline parallelism.

\subsection{Tensor Parallelism}
\begin{figure*}[ht]
    \centering
    \includegraphics[width=0.8\textwidth]{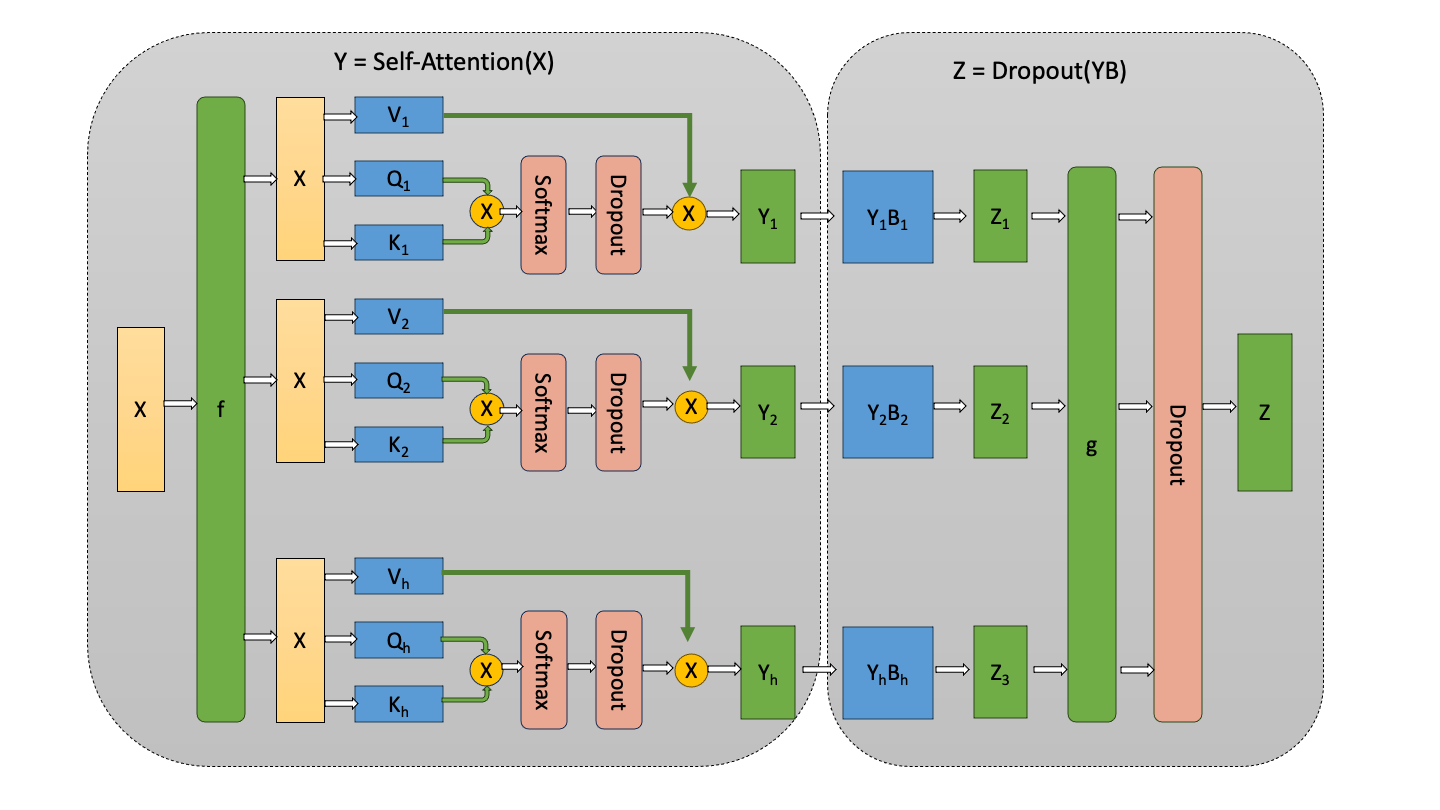}
    \caption{Tensor parallelism on Attention block.}
    \label{fig:tensor-parallelism}
\end{figure*}

Tensor parallelism splits the weight tensor of a layer across the rows. The attention block has three weight tensors $K, Q, V \in \mathbb{R}^{s\times d}$ for a given layer. These tensors are split column-wise (\ref{fig:tensor-parallelism}) and $K = [K_1, K_2, \dots, K_h]$, $Q = [Q_1, Q_2, \dots, Q_h]$, and $V = [V_1, V_2, \dots, V_h]$ where $K_i, Q_i, V_i \in \mathbb{R}^{s\times d/h}$.

\begin{comment}
\begin{figure*}[t!]
    \centering
    \begin{subfigure}[t]{0.5\textwidth}
        \centering
        \includegraphics[height=1.2in]{images/tp-attention.png}
        \caption{\textcolor{red}{[Reproduce] Tensor parallelism for the attention block.}}
    \end{subfigure}%
    ~ 
    \begin{subfigure}[t]{0.5\textwidth}
        \centering
        \includegraphics[height=1.2in]{images/tp-ffn.png}
        \caption{\textcolor{red}{[Reproduce] Tensor parallelism for the FFN block.}}
    \end{subfigure}
    \caption{Tensor Parallelism}
    \label{fig:tensor-parallelism}
\end{figure*}
\end{comment}

For attention computation, the original formula is $softmax(\frac{KQ^T}{\sqrt{d}})V$. For each split, the partial attention is computed as $softmax(\frac{K_iQ_i^T}{\sqrt{d/h}})V_i$. These partial attentions are then appended along column dimension to find the final attention $A = [A_1, A_2, \dots, A_h]$ where $A\in\mathbb{R}^{s\times d}, A_i \in \mathbb{R}^{s \times d/h}$ (Figure~\ref{fig:tensor-parallelism}(b)).

For the FFN block, the computed attention goes through two FFN layers, where $A$ is first multiplied by $W_1 \in \mathbb{R}^{d\times 4d}$, then the result is multiplied by $W_2 \in \mathbb{R}^{4d \times d}$. $W_1$ is partitioned along column dimension, whereas $W_2$ is partitioned along row dimension (\ref{fig:tensor-parallelism}(b)).

\subsection{Pipeline Parallelism}
Pipeline parallelism splits the model into $p$ stages, each having roughly $L/p$ layers. Then, the batch is split into micro-batches, and at every execution step, one micro-batch is passed through a stage. Each stage is placed on a GPU. Initially, only the first GPU can process the first micro-batch. At the second execution step, the first micro-batch progresses to the second stage, and the first micro-batch can now go to the first stage. This is repeated until the last micro-batch reaches the last stage. Then, the backward propagation starts, and the whole process continues in the reverse direction. GPipe proposes this simple method. Synchronization points are introduced after every batch to maintain the correct order of computation, requiring flushing pipeline stages. So, at the beginning and the end of a batch's processing, GPUs hosting earlier and later stages stay idle, resulting in wasted compute time or a \textit{pipeline bubble}. The pipeline bubble fraction is $\frac{p-1}{m}$, where $m$ is the number of micro-batches in a batch.

The simple GPipe~\cite{huang2019gpipe} scheduling creates a large pipeline bubble. Some additional methods are in place to reduce the pipeline bubbles. One such way is 1F1B scheduling proposed by PipeDream~\cite{narayanan2019pipedream}, wherein during the forward pass, initially, the micro-batches are allowed to flow forward until the last group receives the first micro-batch. But then the backward propagation of the first batch starts, and from then, a forward pass is always accompanied by a backward pass, hence the name 1F1B. An interleaving schedule has been proposed to reduce the bubble size even further
where instead of placing one pipeline group on one GPU, multiple smaller pipeline groups are placed on a single GPU.

\begin{comment}
\begin{figure*}[ht]
    \centering
    \includegraphics[width=0.8\textwidth]{images/1F1B-interleaved.png}
    \caption{\textcolor{red}{[Reproduce] 1F1B scheduling with interleaving.}}
    \label{fig:1f1b-interleaving}
\end{figure*}
\end{comment}

The pipeline bubble size for the 1F1B schedule is roughly $p/m$, where p is the number of pipeline groups, and m is the number of micro-batches. For the 1f1B schedule with interleaving, the bubble size is $\frac{p-1}{m\times v}$, where $v$ is the number of interleaved groups placed on a single GPU.

\subsection{Sharded Data Parallelism}
\begin{figure*}[ht]
    \begin{subfigure}[t]{0.45\textwidth}
      \includegraphics[height=2in]{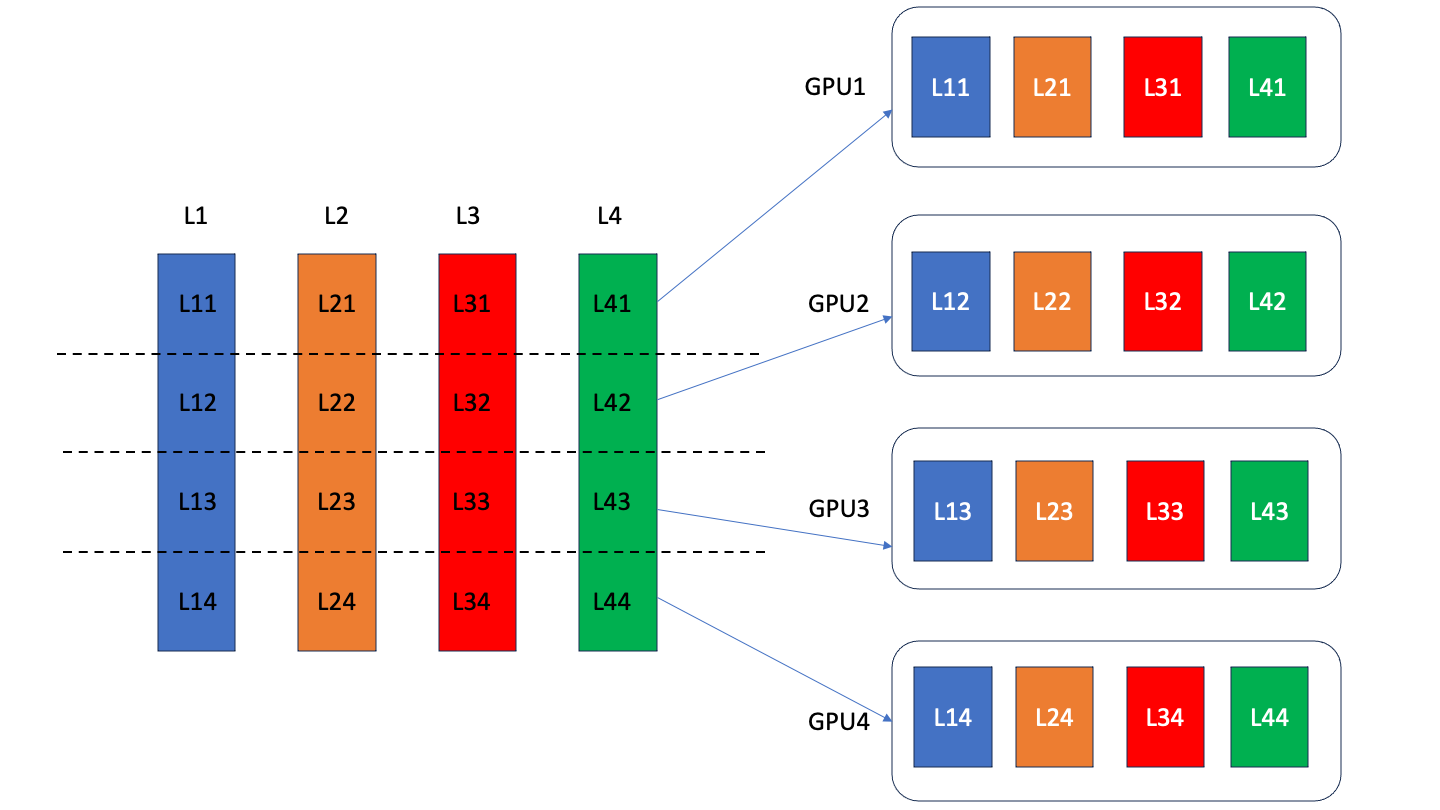}
    \caption{Model is sharded vertically, and each shard is placed on a GPU.}
    \end{subfigure}
    ~
    \begin{subfigure}[t]{0.45\textwidth}
        \includegraphics[height=2in]{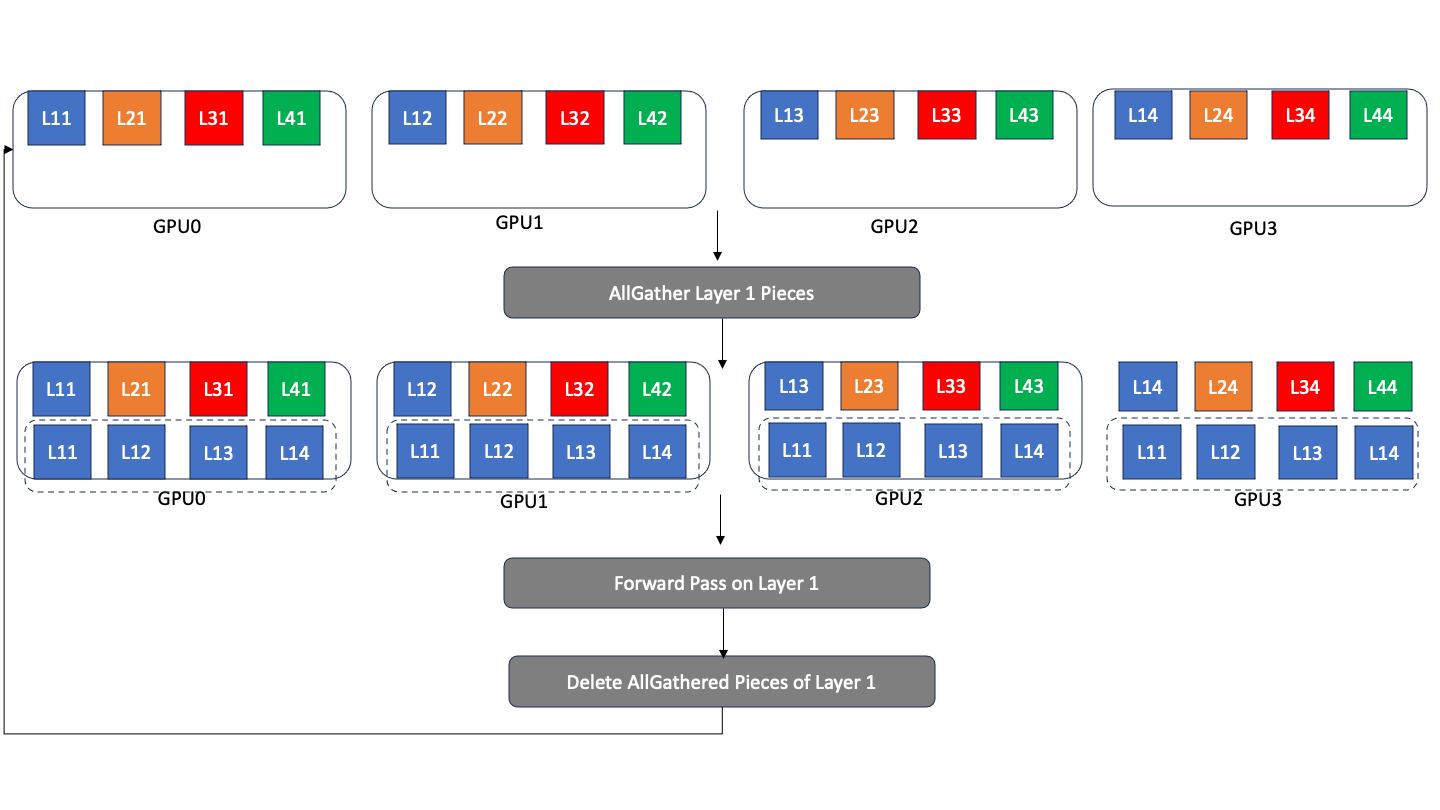}
        \caption{Sharded Data Parallelism, illustrated for the first layer.}
        \label{fig:sharded-data-parallelism}
    \end{subfigure}
    \
\end{figure*}

Sharded data parallelism shards model parameters, optimizer states, and gradients row-wise and places one partition on each GPU. Since training advances one layer at a time, the computing devices need to have only one full layer and associated values (optimizer states, gradients, and parameters) in the memory. Sharded data parallelism takes advantage of this; before execution of a layer, that layer is materialized in all the GPUs by performing all gather across all the GPUs for that layer~\ref{fig:sharded-data-parallelism}. Now, all the GPUs have replicas of the same layer. Then, the layer is executed on different data batches on different GPUs. After that, each GPU deletes all the gathered parts of that layer and prepares for the next layer's materialization through all-gather. This way, it emulates data parallelism, but instead of every GPU hosting a complete replica of the entire model, it just hosts a replica of the currently active layer.

Sharded data parallelism can facilitate data parallel training of a large model across GPUs, even if the model is too large to fit in a single GPU's memory. DeepSpeed's ZeRO optimizers~\cite{rajbhandari2020zero} support sharded data parallelism in varying degrees. ZeRO-1 only shard optimizer states, ZeRO-2 shards gradients along with optimizer states, and ZeRO-3 shards optimizer states, gradients, and model parameters. On the other hand, PyTorch FSDP (Fully Sharded Data Parallelism)~\cite{zhao2023pytorch} shards all three and also supports a hybrid data parallelism by combining sharded data parallelism with traditional data parallelism.

\subsection{3D Parallelism and Megatron-DeepSpeed}
Using only a single parallelism strategy to implement model parallelism can be an inefficient approach. For example, if we use only tensor parallelism to slice the model horizontally, the tensors can be too thin, requiring frequent all-reduce communication that can slow down the training. On the other hand, if we partition the model into too many pipeline stages, each stage will have tiny amounts of computation, which will require frequent communication. A known issue is that performing tensor parallel training across multiple nodes requires slow tree-like allreduce. 

The use of multiple of these modes of parallelism in a hybrid fashion can minimize the areas of poor performance. 3D parallelism combines tensor, pipeline, and data (traditional and sharded) parallelism techniques to utilize resources. Through a proper setup, 3D parallelism can reduce communication latency by overlapping communication with computation. The standard code base for 3D parallelism used across the AI landscape is based on the Megatron-LM~\cite{megatron-lm}. Megatron-DeepSpeed~\cite{megatron-deepspeed} extends on the features from Megatron-LM by adding DeepSpeed features such as the ZeRO-1 sharded data parallelism and a pipeline parallelism with overlapped 1F1B schedule. However, these standard codebases are all developed for NVIDIA GPUs and the CUDA platform.

As the most complete framework, we want to utilize Megatron-DeepSpeed on Frontier, an AMD system whoses software stack is built on the ROCM software platform~\cite{rocm-platform}.

%% file: porting.tex
\subsection{Porting Megatron-DeepSpeed to Frontier}
Megatron-DeepSpeed codebase is forked from NVIDIA's Megatron-LM codebase, and Microsoft then added DeepSpeed ZeRO optimizers, pipeline parallelism, and Mixture of Experts into this. NVIDIA develops Megatron-LM; hence, its codebase is developed with NVIDIA GPUs and CUDA environment as the target platform. Porting this codebase to run on the AMD platform presents some challenges. 

\subsubsection{CUDA Code}
CUDA code doesn't run on AMD hardware; however, HIP, a CUDA-like C/C++ extension language, does. We converted the CUDA source code to HIP code using the \texttt{hipify} tool, built the shareable objects (so files) using \texttt{hipcc}, and then used \texttt{pybind} to make these shareable objects accessible from Python code.

\subsubsection{DeepSpeed Ops}
Most of the DeepSpeed ops are built during the execution of the training pipeline through JIT (Just in time) compilation. However, the JIT compilation of DeepSpeed ops didn't work on the ROCM platform, so we prebuilt all the ops when we installed DeepSpeed. We disabled all JIT functionalities from the Megatron-DeepSpeed codebase to avoid any runtime error.

\subsubsection{Initialization of PyTorch Distributed Environment}
Megatron-DeepSpeed utilized PyTorch Distributed initialization for creating various data and model parallel groups. This initialization process requires dedicating one compute node as the "master" node, and all the distributed processes require its IP address. We modify the codebase to accept $MASTER\_ADDR$ as an argument. We prepared a launch script to read the first node's IP address from the SLURM node list and pass this as an argument to all the processes launched using srun. The initialization code then uses this $MASTER\_ADDR$ for PyTorch Distributed initialization.

\subsubsection{Libraries/Packages provided through ROCM Platform Software}
We worked with AMD developers to get the ROCM version of some of the essential CUDA packages, such as APEX~\cite{nvidia-apex}. APEX is NVIDIA's mixed precision library, which is heavily leveraged by the Megatron-DeepSpeed code base for mixed precision training. We also adapted ROCM-enabled versions of FlashAttention~\cite{dao2022flashattention} and FlashAttention2~\cite{dao2023flashattention} libraries for use with available compilers on Frontier. The Flash-Attention operations are ported to AMD GPUs using kernels from the Composable Kernel library\footnote{\url{https://github.com/ROCm/composable_kernel}}.

%% file: empirical-analysis.tex
\section{Empirical Analysis of Various Distribution Strategy}\label{sec:empirical}
In this section, we report our experiments exploring various distribution strategies and their optimal parameter values (Table~\ref{tab:dist-parameters}). 

\begin{table}[h]
    \centering
    \begin{tabular}{| p{3cm} | p{4cm} |}
        \hline
        Distribution Strategy & Tunable Parameters  \\ \hline \hline 
        Tensor Parallelism & Tensor Parallel Size ($TP$) \\ \hline 
        Pipeline Parallelism & Pipeline Parallel Size ($PP$), \#Microbatches ($m$) \\ \hline 
        Sharded Data Parallelism & ZeRO-1 \\ \hline 
        Common & Micro Batch Size \\ \hline 
        Mixed Precision Training & FP16, BF16 \\ \hline 
    \end{tabular}
    \caption{Distribution Strategies and relevant tunable parameters}
    \label{tab:dist-parameters}
\end{table}

\subsection{Tensor Parallelism}
Tensor parallelism partitions model layers row-wise, and after every layer, the partial activation values need to be aggregated via allreduce. AllReduce after every layer execution is costly, and this depends on communication bandwidth between GPUs in a tensor-parallel group, communication volume which depends on hidden size and micro-batch size.

\begin{figure}[h]
    \centering
    \includegraphics[width=0.5\textwidth]{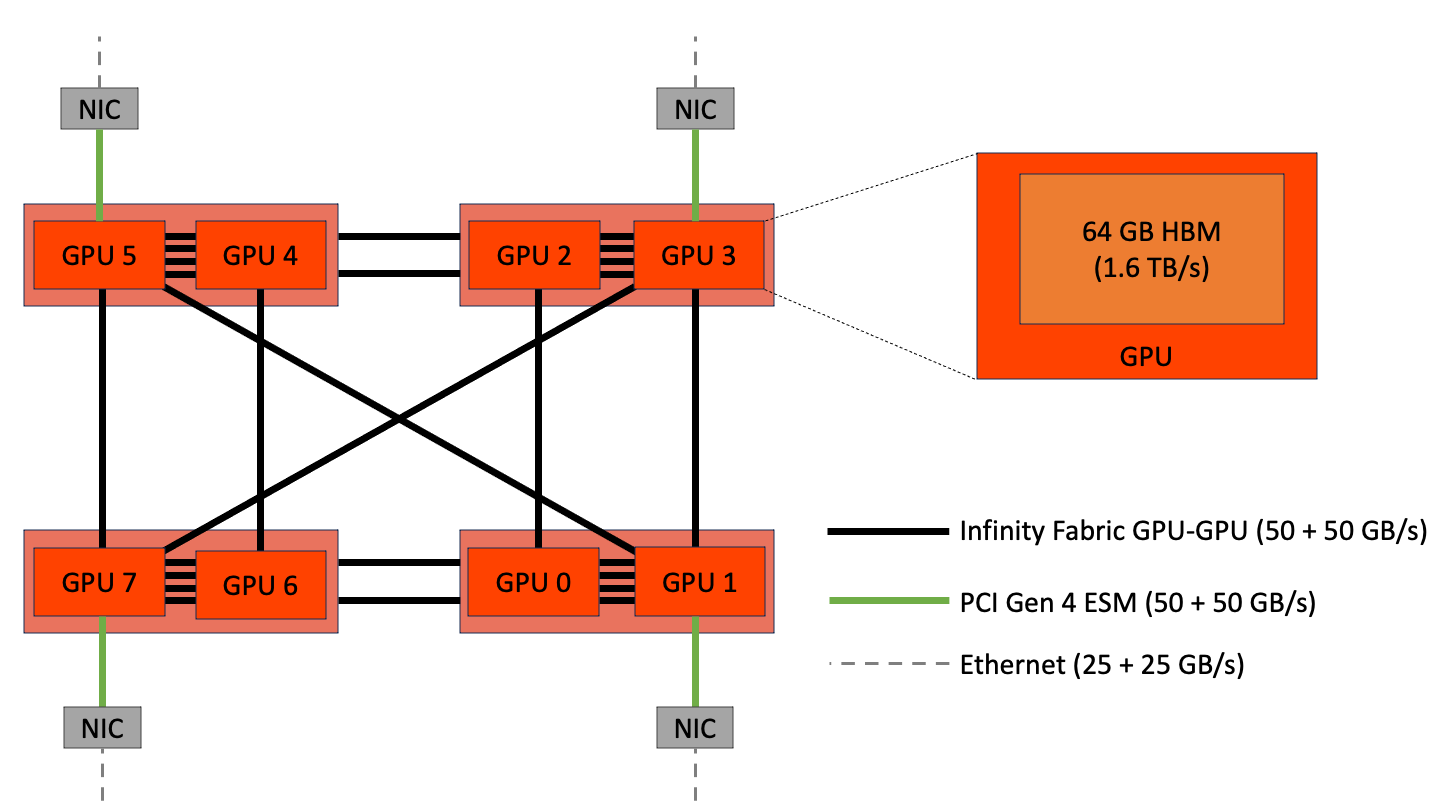}
    \caption{Communication Bandwidth between GPUs in Frontier.}
    \label{fig:frontier-node}
\end{figure}

Figure~\ref{fig:frontier-node} shows the communication bandwidth between Frontier GPUs. There are 8 GPUs in a node, and the GPUs in a single die are connected via four (50+50 GB/s) infinite fabrics. The bandwidth between GPUs across the die is half of it. But, the bandwidth between GPUs across nodes is 25+25 GB/s. So, from the network topology and configuration, TP = 2 would provide the fastest communication, and TP = 4 or 8 would be the second fastest. But, for TP > 8, the communication will happen over slower ethernet, and the communication will be much slower. So, keeping TP in [2, 4, 8] should be the optimal strategy.

\begin{figure}[h]
    \centering
    \includegraphics[width=0.45\textwidth]{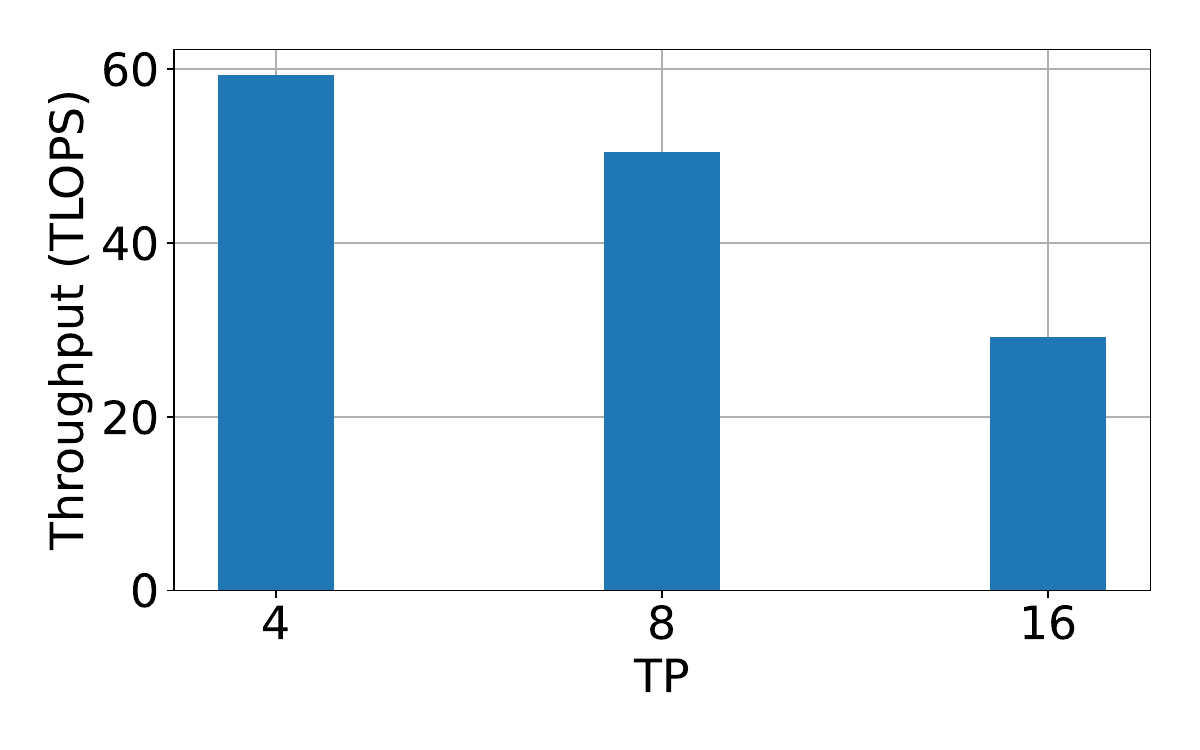}
    
    \caption{GPU throughput vs TP for a 22B model. }
    \label{fig:throughput-vs-tp}
\end{figure}

We train a 1.4B model using 8 GPUs by varying TP from 1 to 8 and see that the smaller the value of TP, the higher the throughput~\ref{fig:throughput-vs-tp}.

\begin{theorem}
Larger values of TP deteriorate training performance.
\end{theorem}

\subsection{Pipeline Parallelism}
Pipeline parallelism partitions the model along the layer dimension and groups consecutive layers into pipeline stages. The execution of a micro-batch flows from one stage to the next one.

A pipeline bubble would be a limiting factor for efficient training with this parallelism. The bubble size is roughly $\frac{TP}{v\times M}$, where TP is the number of pipeline stages, M is the number of micro-batches, and v is the number of overlapped pipeline stages on a single GPU. A large M can ensure the bubble size is minimal; however, it might need to utilize gradient accumulation. A large M results in a sizeable global batch size (GBS).

\begin{figure*}[h]
\centering
    \begin{subfigure}[t]{0.45\textwidth}
        \includegraphics[width=1.0\textwidth]{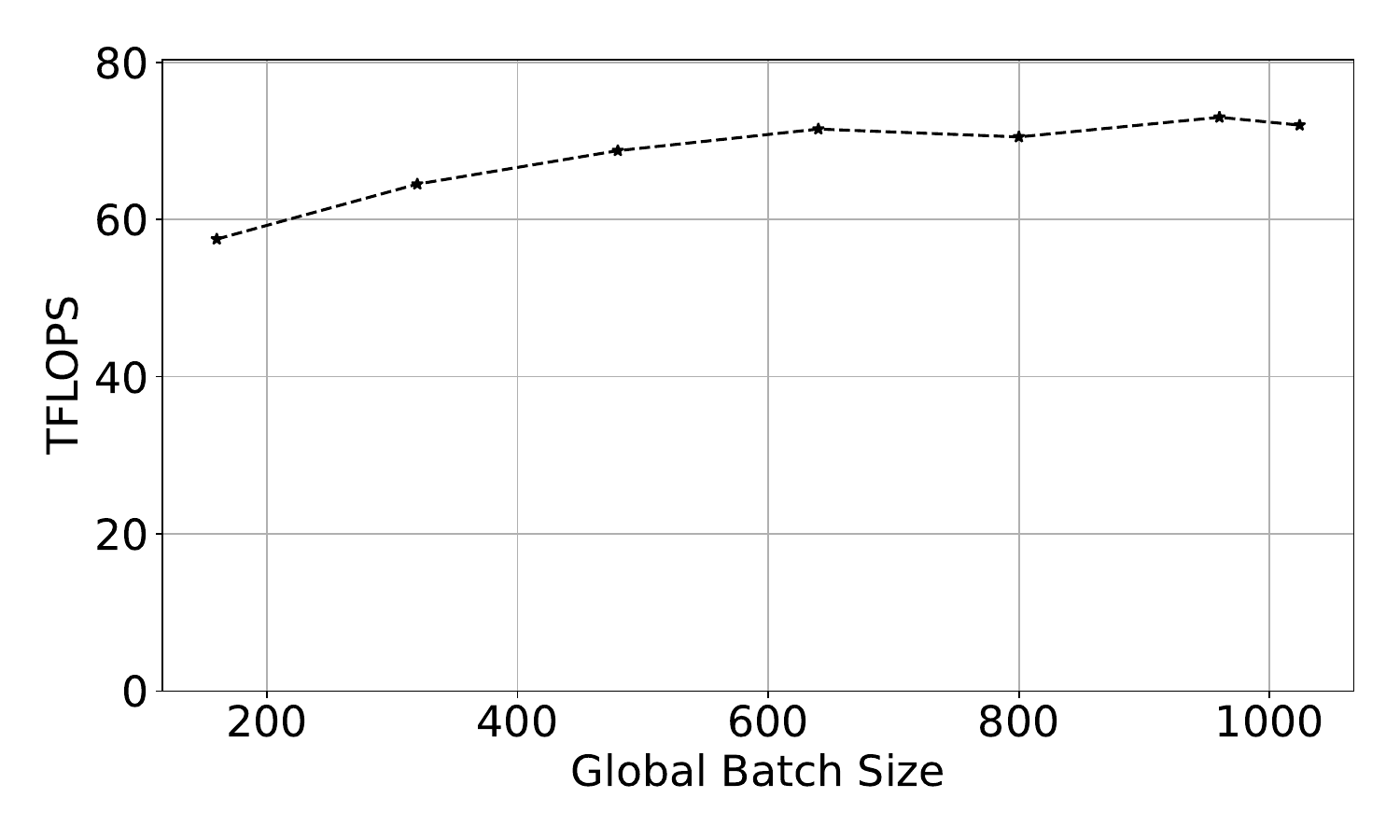}
        \caption{Throughput vs. global batch-size for 22B model.}
        %\label{fig:tflops-vs-gbs-22b}
    \end{subfigure}
    ~
    \begin{subfigure}[t]{0.45\textwidth}
        \includegraphics[width=1.0\textwidth]{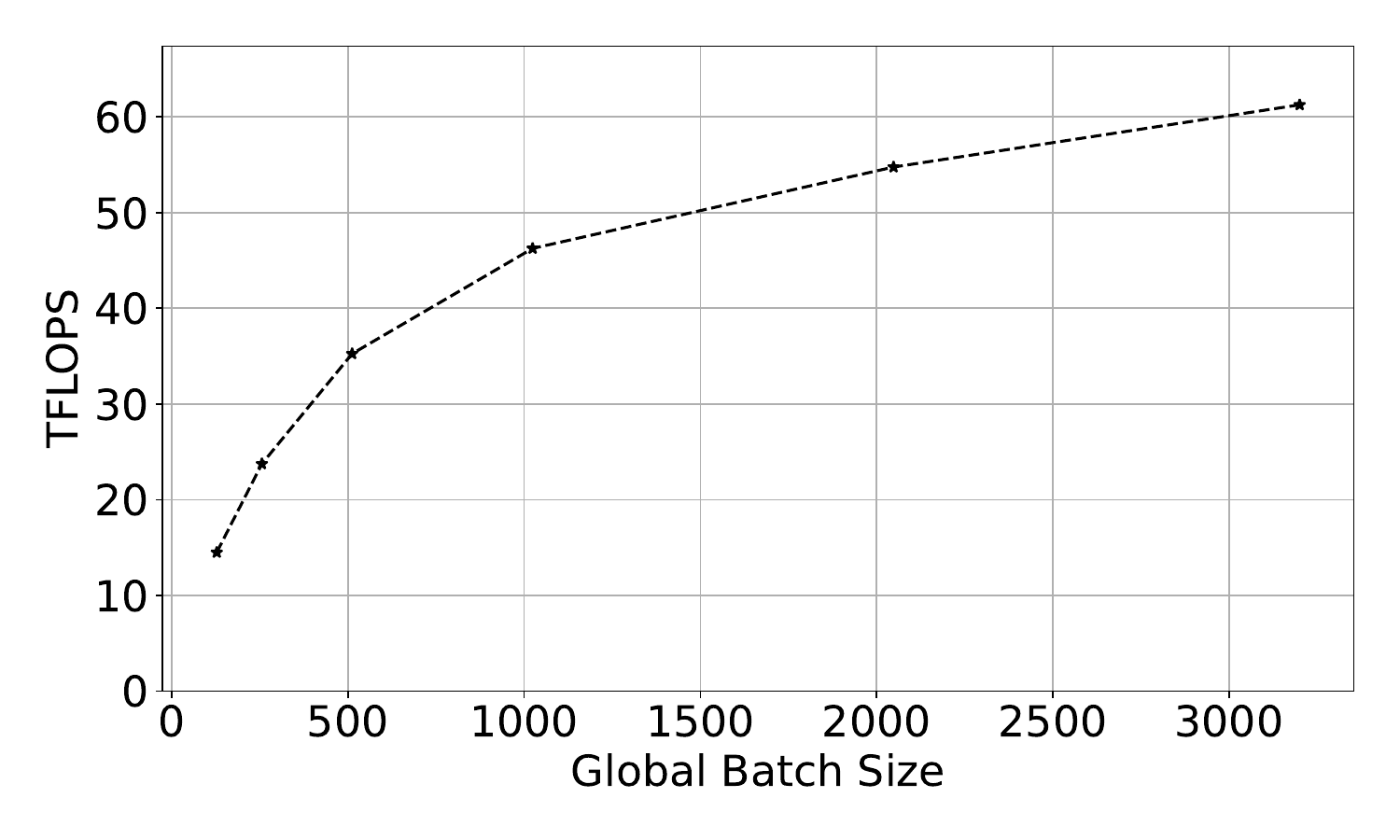}
        \caption{Throughput vs global batch-size for 1T model.}
        %\label{fig:tflops-vs-gbs-1T}
    \end{subfigure}
    \caption{GPU throughput as a function of global batch size. Throughput increases with global batch size since it increases the number of micro-batches (M) and reduces the bubble size. However, using a large global batch size for a single replica has implications for how much data parallelism we can use and what part of the entire machine we will be able to use.}
    \label{fig:throughput-vs-batch-size}
\end{figure*}

We saw the effect of large M or large GBS to see the impact on GPU throughput for two models of size 22B parameters and 1T parameters (Figure~\ref{fig:throughput-vs-batch-size}).

\begin{theorem}
    Saturating pipeline stages with large global batch sizes or many micro-batches minimizes pipeline bubble size.
\end{theorem}

\subsubsection{Impact of Number of Pipeline Stages}
Next, we investigate the impact of the number of pipeline stages on training performance. Intuitively, more pipeline stages mean less computation before the communication happens. With a fixed global batch size (number of micro-batches), the pipeline bubble size increases with the number of pipeline stages. We also experiment with increasing the number of pipeline stages while keeping the $\frac{PP}{M}$ fixed by increasing the global batch size proportionally.

\begin{figure*}[h]
    \centering
    \begin{subfigure}[t]{0.45\textwidth}
        \includegraphics[width=1.0\textwidth]{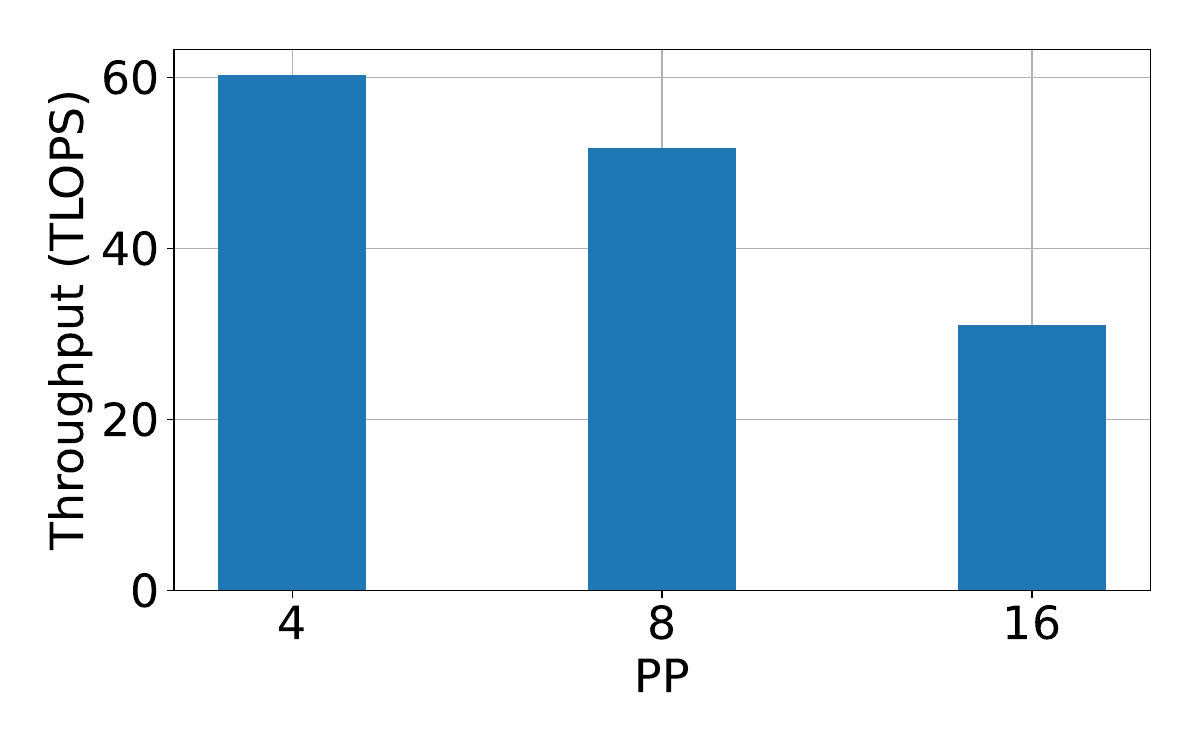}
    \caption{Throughput vs. PP while keeping global batch size fixed at 128.}
    \label{fig:throughput-vs-pp-gbs-128}
    \end{subfigure}
    ~
    \begin{subfigure}[t]{0.45\textwidth}
        \includegraphics[width=1.0\textwidth]{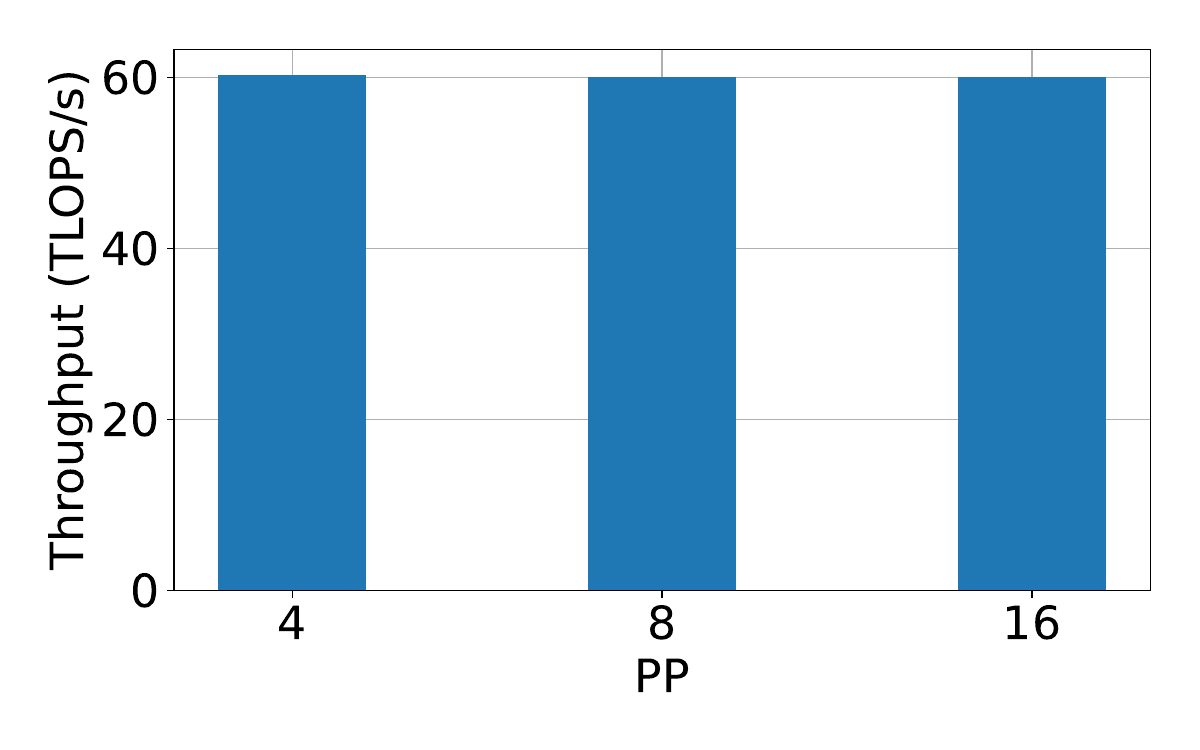}
    \caption{Throughput vs. PP while scaling global batch size to keep the pipeline bubble ratio fixed.}
    \label{fig:throughput-vs-pp-scaled-gbs}
    \end{subfigure}
    \label{fig:throughput-vs-pp}
    \caption{Impact of pipeline parallelism on training performance.}
\end{figure*}

\begin{theorem}
    Increasing the number of pipeline stages while keeping the global batch size fixed increases the pipeline bubble size and deteriorates training performance.
\end{theorem}

\begin{theorem}
    The training performance can be maintained with an increasing number of pipeline stages if the ratio of the number of pipeline stages to the number of micro-batches is kept constant.
\end{theorem}

From the first experiment (Figure~\ref{fig:throughput-vs-pp-gbs-128}), the training performance deteriorates with the increasing pipeline stages. However, scaling global batch size to fix the bubble ratio maintains the throughput (Figure~\ref{fig:throughput-vs-pp-scaled-gbs}).

%% file: deephyper.tex
\section{Hyperparameter Tuning Using DeepHyper}\label{sec:deephyper}

DeepHyper is a Python package developed to automate the development of machine learning workflows with algorithms such as hyperparameter optimization (HPO)~\cite{balaprakash2018deephyper}, neural architecture search~\cite{maulik2020recurrent}, and uncertainty quantification~\cite{egele2022autodeuq}. Hyperparameters are the parameters of the optimized learning workflow that define it but cannot be inferred during the so-called ``training'' phase.

To find the best distributed training strategy determined by tensor, data pipelining, data sharding, and data parallelism settings, we use the asynchronous HPO from DeepHyper, which is based on a Bayesian optimization solver.

For the execution aspect, DeepHyper offers multiple parallelization schemes~\cite{egele2023asynchronous} (e.g., centralized or decentralized for the search execution; threads, processes, or MPI for the execution of our black box). We focused on a centralized architecture with processes for black-box evaluations, as our setting did not reach any parallelism bottleneck. Then, as each evaluation also requires a set of parallel resources (in particular, different nodes), we pass a queue of all nodes available to DeepHyper and ask each job to use 16 nodes. When DeepHyper suggests a hyperparameter configuration to be evaluated, a local process is created to schedule its evaluation; within this process, we set up the distributed training application given the passed hyperparameters and launch it through SLURM via \texttt{srun}. %We provide a flow diagram of the execution in Figure~\ref{fig:deephyper-workflow}{\color{red}Romain: diagram should be updated}.

\begin{comment}
\begin{figure}[h]
    \centering
    \includegraphics[width=0.5\textwidth]{images/DeepHyper.png}
    \caption{DeepHyper workflow.}
    \label{fig:deephyper-workflow}
\end{figure}
\end{comment}

We define the space of allowed hyperparameters through DeepHyper, which can be of categorical, discrete, or continuous types (a.k.a. mixed types). Table~\ref{tab:tuning-175B} describes the hyperparameter space we used to tune the distributed training of a 175 Billion parameter model through FLOPs maximization. %{\color{red}Romain: I advise, to avoid confusion, to use the word ``parameter'' for the model's learnable weights and ``hyperparameters'' for variables optimized by deephyper} 
In this space, hyperparameter configurations can exist, triggering failures of the distributed training, such as memory exhaustion (out-of-memory). We handle such failures by catching the exception and returning the special \texttt{F}-objective value to DeepHyper, which internally penalizes such evaluations to discourage future evaluations.

\begin{table}[h]
    \centering
    \begin{tabular}{c|c}
    \hline
    Hyperparameters & Range  \\ \hline \hline
    Pipeline-parallel-size (PP) & $PP \in \{1, 2, 4, 8, 12, 16\}$ \\ \hline
    Tensor-parallel-size (TP) & $TP \in \{1, 2, 4, 8\}$ \\ \hline 
    Micro-batch-size (MBS) & $MBS \in [4, 20]$ \\ \hline
    Gradient accumulation steps (GAS) & $GAS \in \{5, 10\}$ \\ \hline 
    ZeRO-1 Optimizer & $ZeRO-1 \in \{True, False\}$ \\ \hline 
    Number of Nodes (NNODES) & $NNODES \in \{12, 16\}$ \\ \hline 
    \end{tabular}
    \caption{Hyperparameter Tuning for 175B Model}
    \label{tab:tuning-175B}
\end{table}

%{\color{red}Romain: describe the experimental setting in a paragraph here}
We ran these hyperparameter tunings on 128 nodes, and each search job utilized 16 nodes. Each job picks 16 nodes from the queue. However, some of the jobs will use 12 nodes or 16 nodes. We dynamically create a srun launch command with these hyperparameters and submit the job for a maximum of 20 minutes.

\begin{figure}
    \centering
    \includegraphics[width=0.5\textwidth]{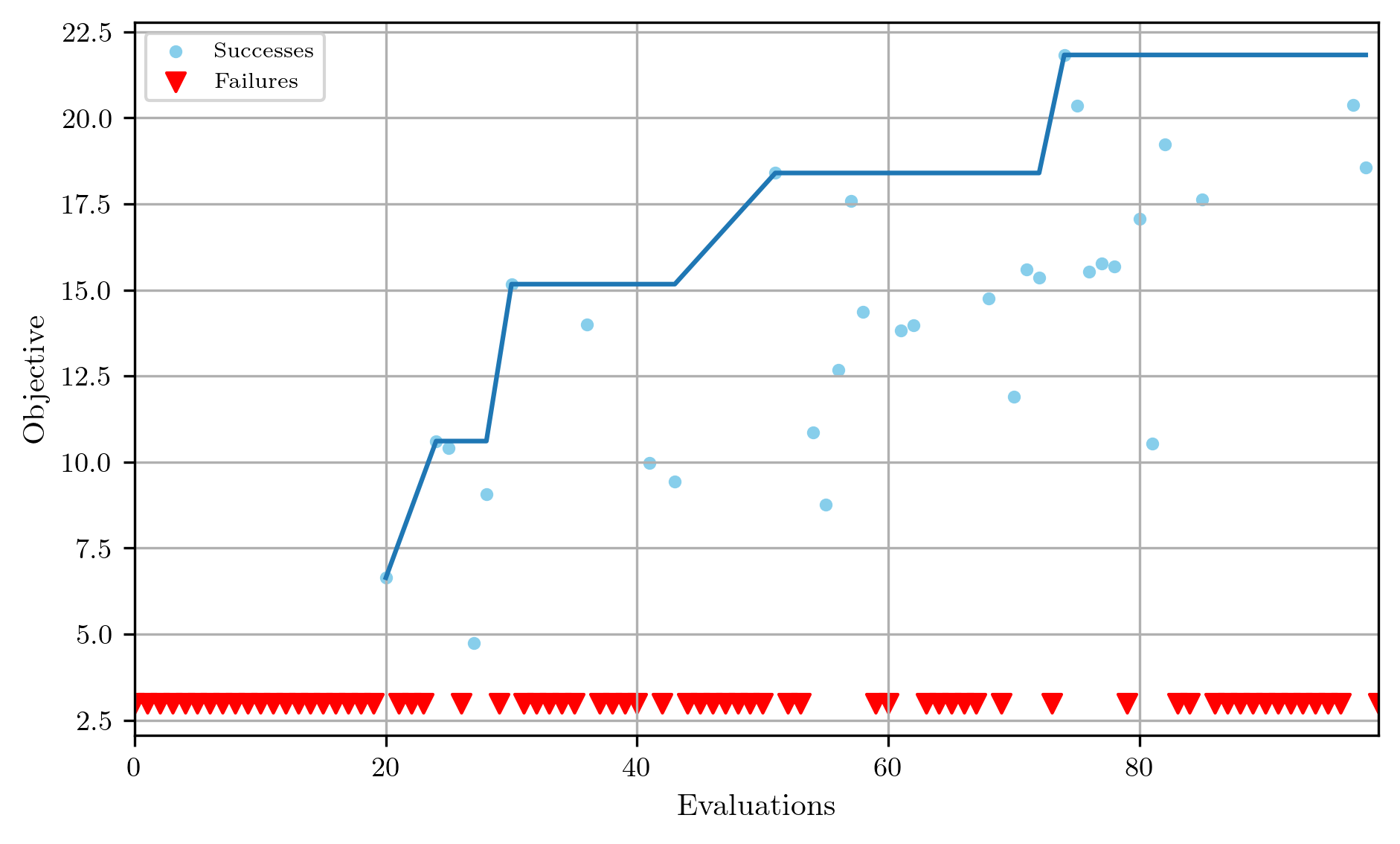}
    \caption{Search Trajectory of DeepHyper search}
    \label{fig:deephyper-search-trajectory}
\end{figure}

In Figure~\ref{fig:deephyper-search-trajectory}, we present the results of our experiment. We observe many failures (red arrows), mostly out-of-memory errors. However, we observe that the frequency of such failures decreases with time.
%{\color{red}Romain: add numbers of graph}. 
For successful evaluations, the best value of FLOPs improves with time to reach a final value of 22 FLOPS. %{\color{red}Romain: replace X}

\begin{comment}
\begin{table}[]
    \centering
    \begin{tabular}{c|c}
    \hline
    Distribution Parameter & Search-Space \\ \hline \hline
    Pipeline-parallel-size (PP) & $PP \in \{32, 64, 128\}$ \\ \hline
    Tensor-parallel-size (TP) & $TP \in \{4, 8, 16\}$ \\ \hline 
    Micro-batch-size (MBS) & $MBS \in [1, 16]$ \\ \hline
    Gradient accumulation steps (GAS) & $GAS \in \{10\}$ \\ \hline 
    ZeRO-1 Optimizer & $ZeRO-1 \in \{True, False\}$ \\ \hline 
    Number of Nodes (NNODES) & $NNODES \in \{128\}$ \\ \hline 
    \end{tabular}
    \caption{Tuning 1T Model}
    \label{tab:tuning-1T}
\end{table}
\end{comment}

\begin{figure}[h]
    \centering
    \includegraphics[width=0.5\textwidth]{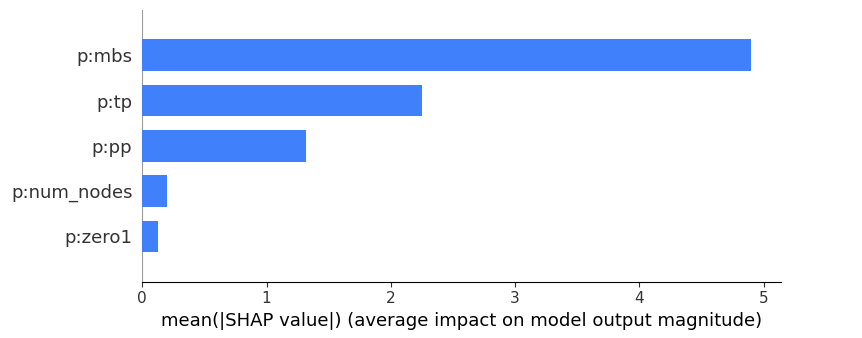}
    \caption{Impact of various hyperparameters on training performance in terms of GPU throughput.}
    \label{fig:sensitivity}
\end{figure}

Using tuning run results of the DeepHyper search, we conduct SHAP (SHapley Additive exPlanations)~\cite{NIPS2017_7062} sensitivity analysis to assess the impact of the hyperparameters on the performance. SHAP is a game theory-based approach for explaining the output of machine learning models. It assigns each feature an importance value for a particular prediction. These values are derived from the Shapley values in cooperative game theory and provide a measure of the contribution of each feature to the prediction. SHAP values are instrumental because they are consistent and locally accurate, meaning they sum up the difference between the model and baseline outputs.

In the context of hyperparameter sensitivity analysis, SHAP can be used to evaluate the impact of different hyperparameters on performance by fitting a regression model that predicts the performance (outputs) with hyperparameters (inputs). In this analysis, various hyperparameters, such as micro-batch size (p:mbs), tensor-parallel-size (p:tp), pipeline-parallel-size (p:pp), number of nodes (p:num\_nodes), and another parameter (p:zero1), are assessed to determine their influence on the FLOPS. When varying these hyperparameters, the SHAP values indicate the average impact on the model output magnitude. The bar chart in Figure~\ref{fig:sensitivity} presents these hyperparameters' mean absolute SHAP values, providing insight into which parameters have the most substantial effect on the model's computational efficiency. 

From the sensitivity analysis of all the hyperparameters (Figure~\ref{fig:sensitivity}), we find that the micro-batch size is the most impactful hyperparameter. Then, tensor-parallel-size, and pipeline-parallel-size. We see that utilizing ZeRO-1 has the least impact.

%% file: recipe.tex
\section{Training a Trillion Parameter Model}\label{sec:training}
From the experiments, hyperparameter tuning, and analysis, we have identified an efficient strategy for training a Trillion-parameter model on Frontier by combining various distribution strategies and software optimizations. In this subsection, we list them with their contribution and best configurations (Table~\ref{tab:best-params-1T-175b}).

\subsection{An Efficient Strategy for Training a Trillion Parameter Model}

\paragraph*{Saturate Pipeline Stages by Increasing the Number of Micro-batches}

We use pipeline parallelism provided by DeepSpeed (from DeepSpeed-Megatron, but not the Megatron's version). This pipeline parallelism algorithm is PipeDream's algorithm, where multiple stages are overlapped, and the 1F1B algorithm is followed to reduce the bubble size. However, the bubble size will increase if the pipeline stages are not saturated. To ensure saturation, the number of micro-batches must equal or exceed the number of pipeline stages.

%\subsubsection{Activation Checkpointing to Increase Per GPU Batch-size}
%To achieve high per-GPU compute utilization, we must facilitate a large micro-batch size. However, a large batch creates a large activation memory requirement, and the program runs into an OOM error. To overcome this, we deploy activation checkpointing, which, at the cost of 30\% extra computation, provides much higher compute utilization.

\paragraph*{Limit Tensor-Parallelism to A Single Node / Eight GPUs}

Since the AllReduce operation is too frequent and needs to be performed for every layer, a layer spread across nodes causes tree-based AllReduction between GPUs across nodes, and the communication latency becomes a significant bottleneck.

\paragraph*{Use Flash-Attention v2}
We observed up to $30\%$ throughput improvement using Flash-attention compared to the regular attention implementation. 

\paragraph*{Use ZeRO-1 Optimizer For Data Parallelism}
We use ZeRO-1 for data parallelism to reduce memory overhead.

\paragraph*{Use RCCL Plugin by AWS to Improve Communication Stability}
AWS OFI RCCL plugin enables EC2 developers to use libfabric as a network provider while running AMD's RCCL-based applications. On Frontier, usage of this plugin shows communication stability.

%and performance improvement of up to $10\%$.

\subsection{Training Performance of a Trillion Parameter Model}
\begin{figure}
    \centering
    \includegraphics[width=0.5\textwidth]{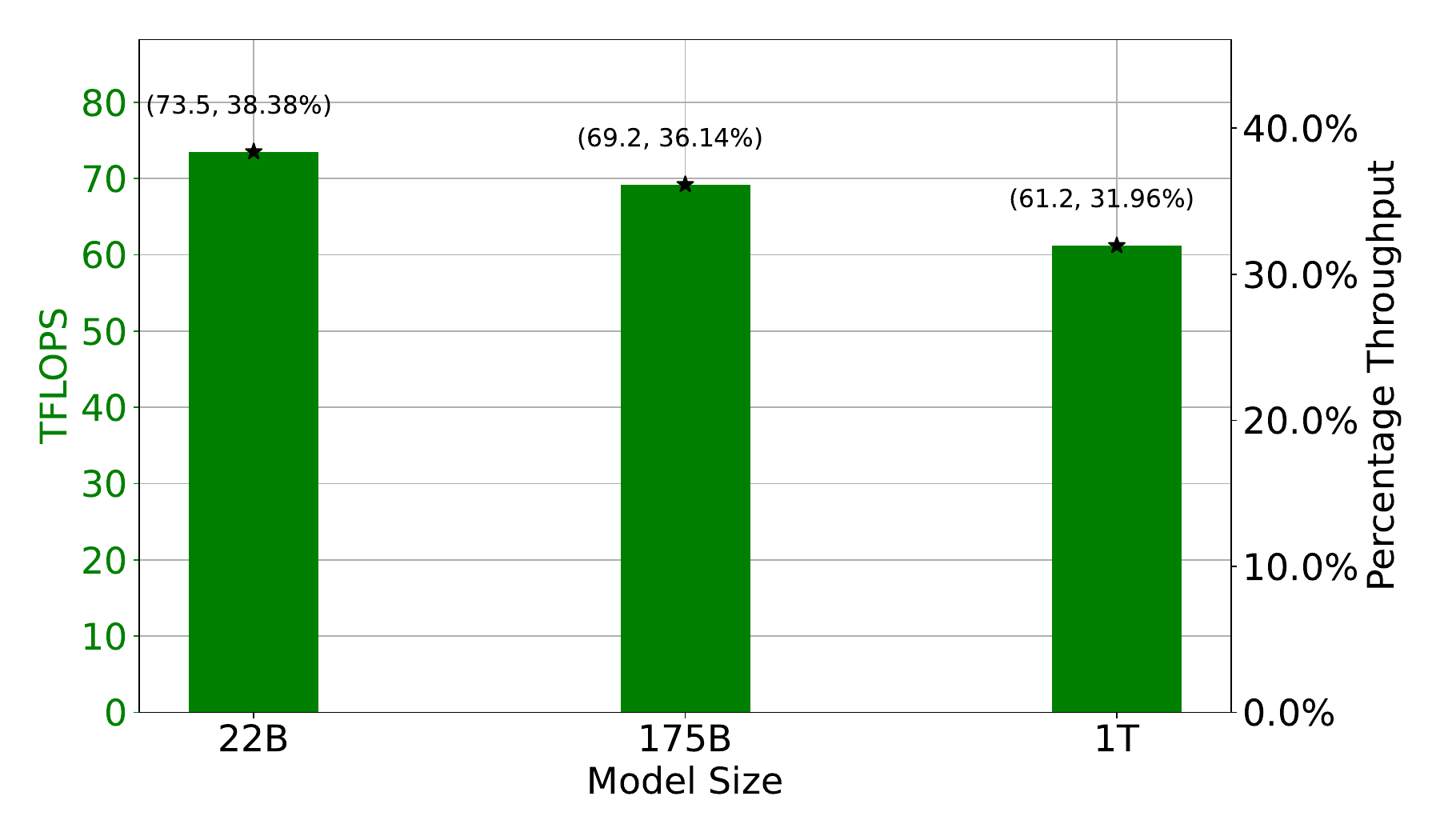}
    \caption{MI250X Throughput for various model sizes. We report the hardware FLOPS, which are in close agreement with the model FLOPS.}
    \label{fig:mi250x-throughput}
\end{figure}

From the lessons learned from hyperparameter tuning, we identified a set of combinations for models of size 22 Billion parameters and 175 Billion parameters. Encouraged by the GPU throughput of these two models, we finally trained a trillion parameter model using the combination of distribution strategies listed in Table~\ref{tab:best-params-1T-175b} for ten iterations to see the training performance. For the 22B parameter model, we could extract $38.38\%$ (73.5 TFLOPS) of its peak throughput (191.5 TFLOPS). For the 175B model training, we achieved $36.14\%$ (69.2 TFLOPs) of peak throughput. Finally, for the 1T model, we achieved $31.96\%$ (61.2 TFLOPs) of peak throughput ~\ref{fig:mi250x-throughput}.

\begin{table}[h]
    \centering
    \begin{tabular}{| c | c | c |}
    \hline 
         Hyperparameters  & \multicolumn{2}{ c |}{Value} \\ \hline 
           &  175B Model & 1T Model \\ \hline \hline 
       TP  & 4 & 8\\ \hline 
       PP & 16 & 64\\ \hline 
       MBS & 1 & 1\\ \hline 
       GBS & 640 & 1600 \\ \hline 
       ZeRO Stage & 1 & 1 \\ \hline 
       Flash Attention & v2 & v2 \\ \hline 
       Precision & fp16 & fp16\\ \hline 
       checkpoint-activations & True & True \\ \hline 
    \end{tabular}
    \caption{Best parameters for training a 175B model and a 1T model.}
    \label{tab:best-params-1T-175b}
\end{table}

\paragraph{Composite Roofline Analysis}
We collected the total hardware flops and bytes read and written throughout the training. From there, we computed the arithmetic intensities for training 22B and 175B parameter models to understand the limiting factors of the kernels. For these models, our achieved FLOPS were 38.38\% and 36.14\%, and arithmetic intensities of 180+. The memory bandwidth roof and the peak flops-rate roof meet close to the arithmetic intensity of 1. Hence, our training is not memory-bound.

\subsection{Scaling Performance}

Sustaining the performance of model-parallel training through data parallelism to engage a large number of GPUs in a system is a challenging task. Frontier GPUs are connected via communication links of various speeds, and stressing the larger part of the network can result in lost performance. So, we scale the training up to 1024 GPUs for a 175B model and 3072 GPUs for a 1T model through data parallelism to measure the scaling efficiency of our training strategy. 

\subsubsection{Weak Scaling}

\begin{figure*}[ht]

    \begin{subfigure}[t]{0.5\textwidth}
        \centering
        \includegraphics[width=1.0\textwidth]{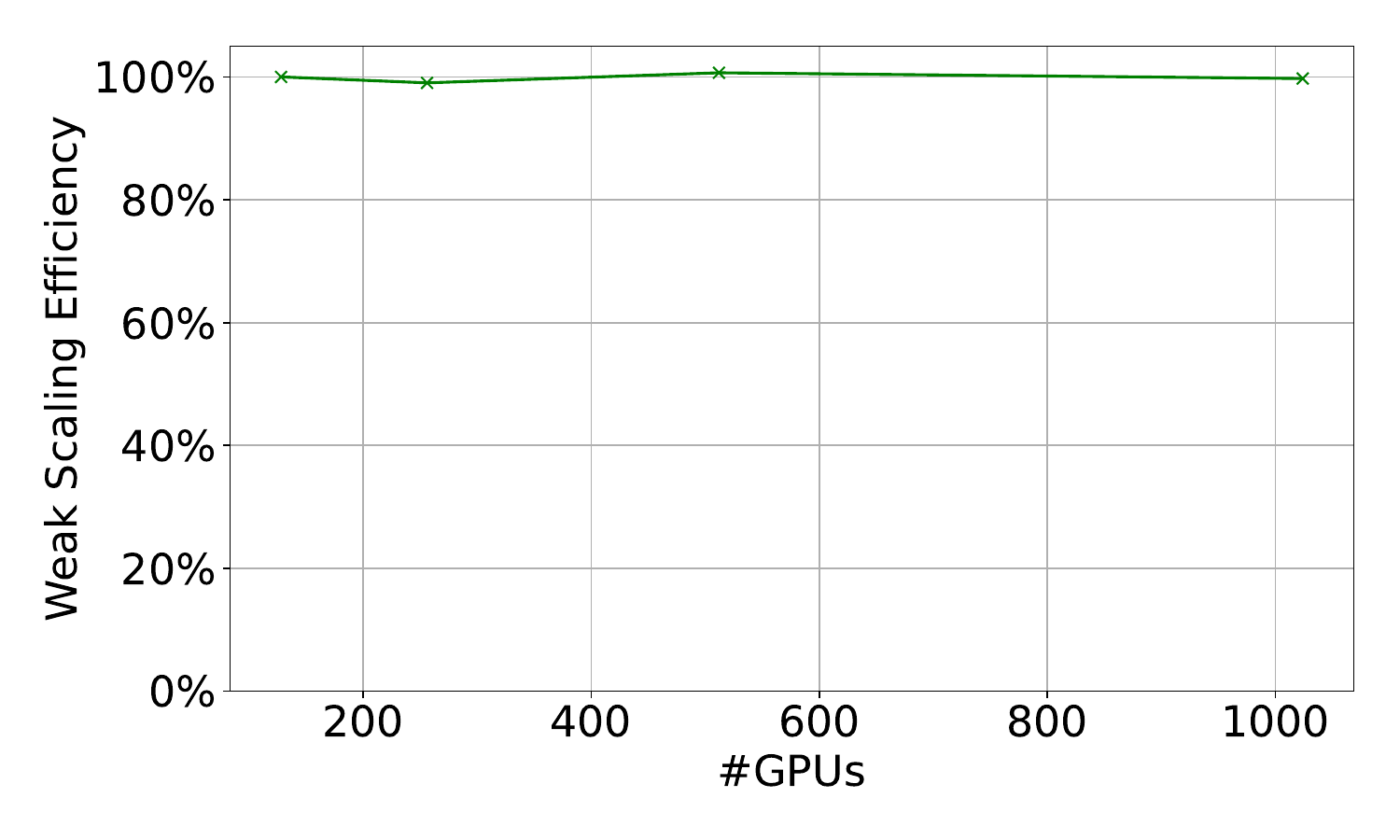}
        \caption{Weak scaling of 175b model training by keeping per replica batch-size fixed at 640.}
        \label{fig:weak-scaling-175b}
    \end{subfigure}
    ~
    \begin{subfigure}[t]{0.5\textwidth}
        \centering
        \includegraphics[width=1.0\textwidth]{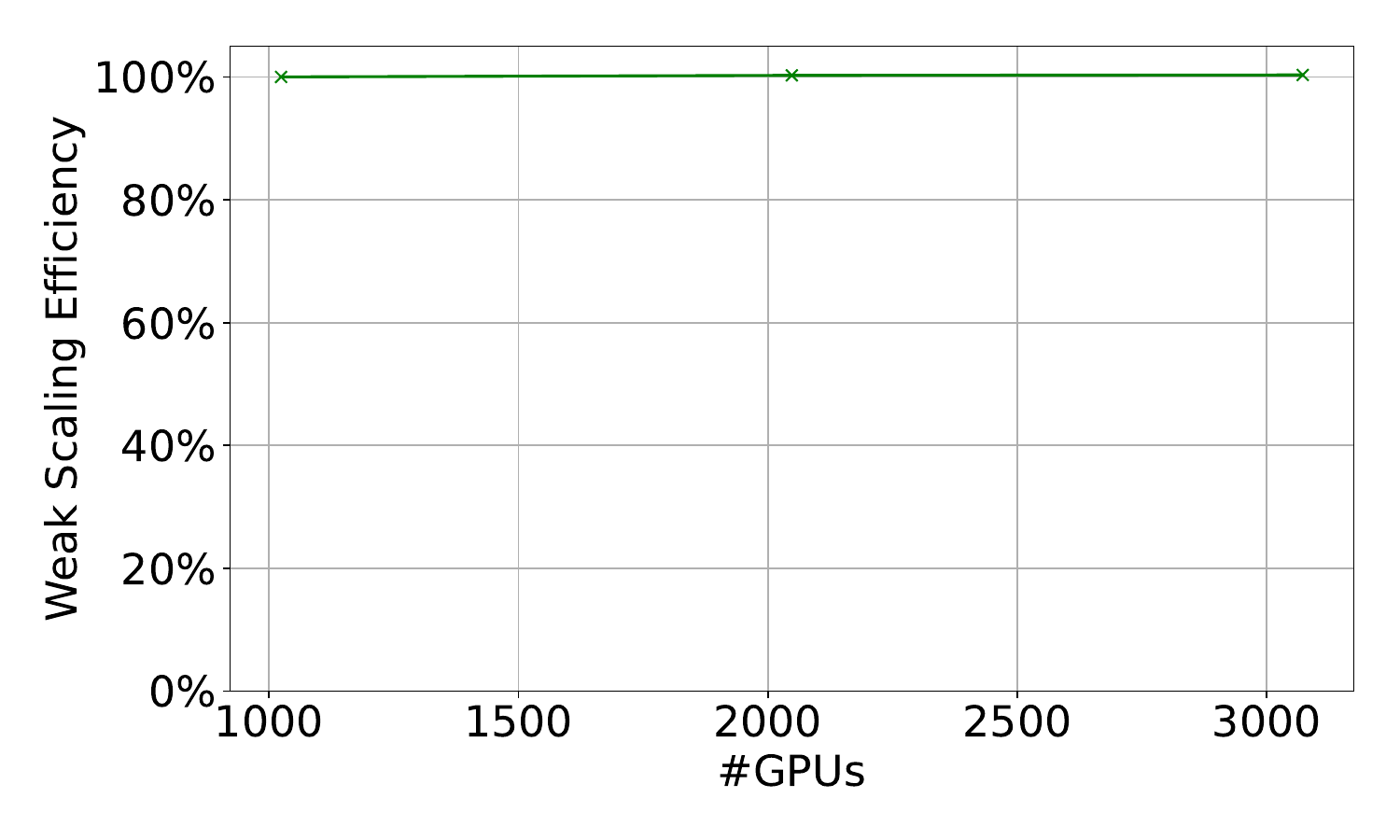}
        \caption{Weak scaling of 1T model training by keeping per replica batch-size fixed at 1600.}
        \label{fig:weak-scaling-1T}
    \end{subfigure}
\caption{Weak scaling performance of 175b model and 1T model training.}
\label{fig:weak-scaling}
\end{figure*}

We perform a weak scaling experiment for the 1T model by performing data-parallel training on 1024, 2048, and 3072 GPUs using global batch sizes 3200, 6400, and 9600. The data-parallel training achieves 100\% weak scaling efficiency (Figure~\ref{fig:weak-scaling}).

\subsubsection{Strong Scaling}
\begin{figure*}[h]

    \begin{subfigure}[t]{0.5\textwidth}
        \centering
        \includegraphics[width=1.0\textwidth]{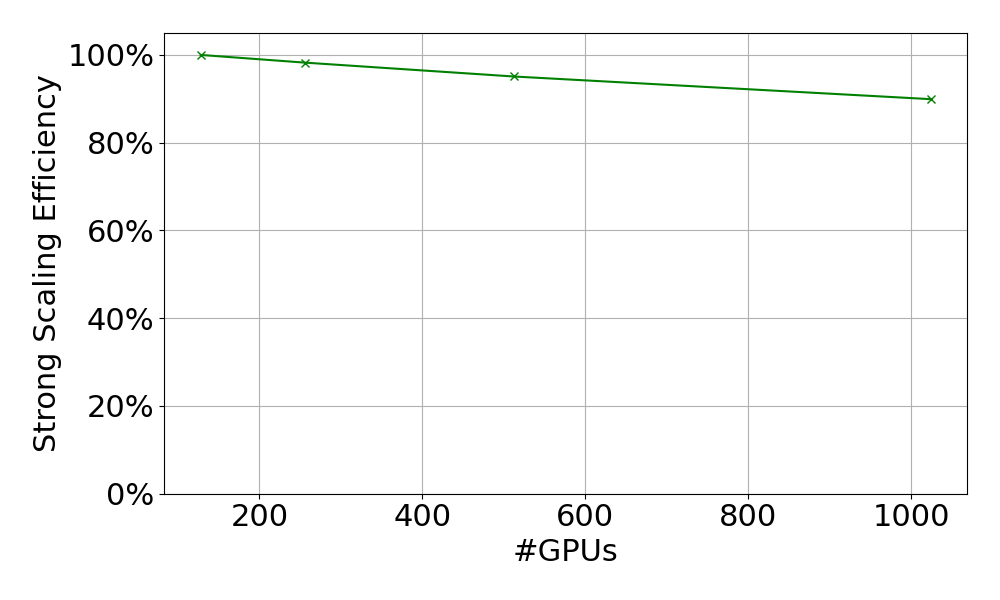}
        \caption{Strong scaling of 175b model training by keeping the total batch size fixed at 8000. The strong scaling efficiency at 1024 GPUs is 89.93\%.}
        \label{fig:strong-scaling-175b}
    \end{subfigure}
    ~
    \begin{subfigure}[t]{0.5\textwidth}
        \centering
        \includegraphics[width=1.0\textwidth]{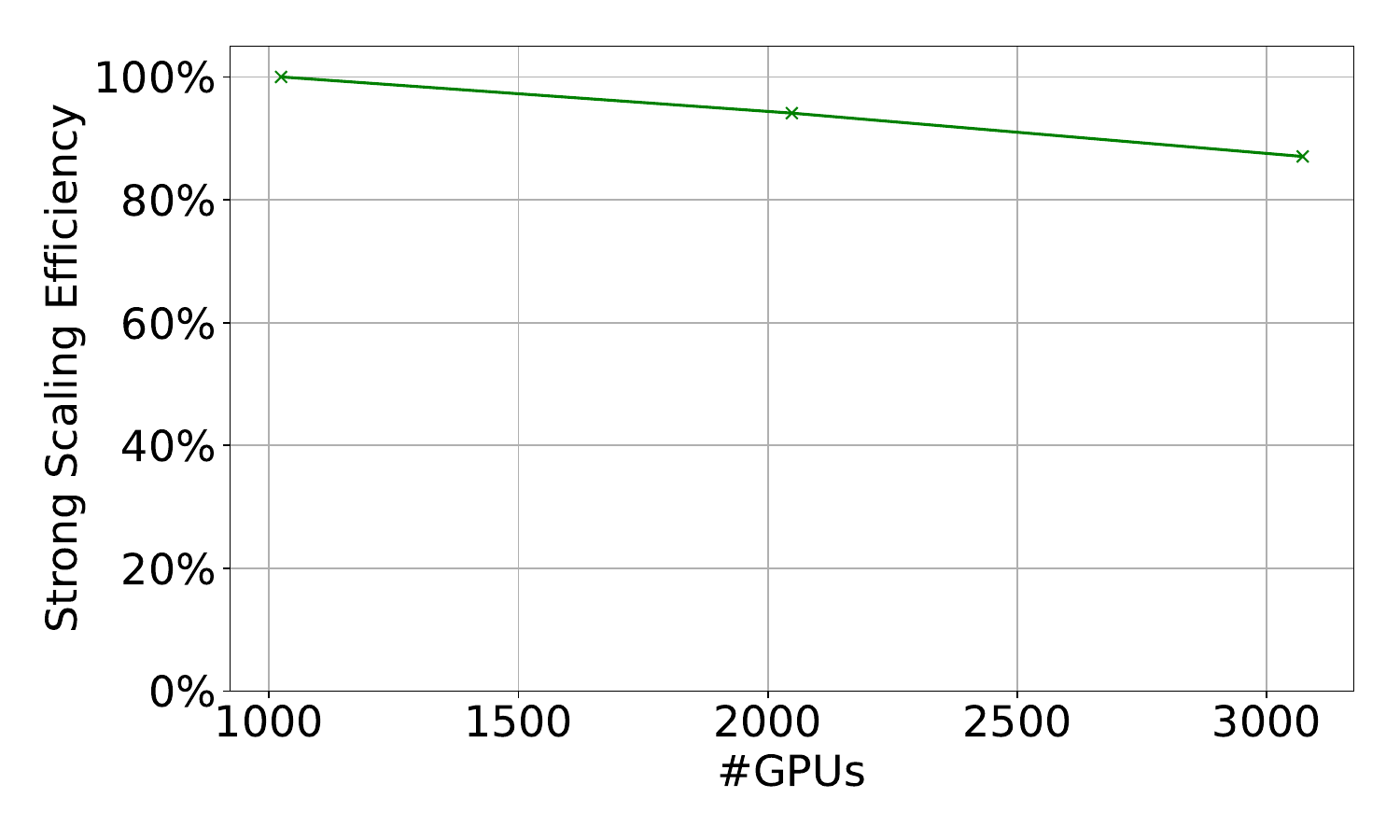}
        \caption{Strong scaling of 1T model training by keeping a total batch-size fixed at 8016. The strong scaling efficiency at 3072 GPUs is 87.05\%.}
        \label{fig:strong-scaling-1T}
    \end{subfigure}
\caption{Strong scaling performance of 175b model and 1T model training.}
\label{fig:strong-scaling}
\end{figure*}

We perform strong scaling experiments by keeping the global batch size at 8000 and then varying the number of GPUs. We achieved $89.93\%$ strong scaling performance for a 175B model on 1024 GPUs (Figure~\ref{fig:strong-scaling-175b}). We achieved $87.05\%$ strong scaling performance for a 1 Trillion parameter model on 3072 GPUs (Figure~\ref{fig:strong-scaling-1T}).

%% file: conclusion.tex
\section{Conclusion and Discussion}
Training an LLM with Billions to Trillions of parameters is a challenging task since we have to orchestrate overcoming the GPU memory wall, minimizing communication latency, and building a software stack with state-of-the-art distribution algorithms. A Trillion parameter model requires a minimum of 14 Terabytes of memory, while an MI250X GPU has only 64 Gigabytes. So, to overcome the memory wall problem, we have explored a combination of model parallelism strategies. However, model parallel training requires communication within groups of GPUs either sharing the parts of the same tensor (Tensor parallelism) or the groups of GPUs hosting neighboring components (pipeline parallelism). We also needed to utilize data parallelism to consume a large number of tokens simultaneously and use a larger number of computing resources to achieve faster time to solution. All of these model parallelism and data parallelism incur communication at various frequencies and of various volumes. The communication latency can significantly increase the training time and reduce the training performance regarding GPU throughput. 

We must select the right combinations of these parallelization and distributed techniques to overlap computation and communication to hide or minimize the latency. We prepared a software stack on Frontier for training LLM models by porting state-of-the-art distributed training frameworks such as Megatron-DeepSpeed and FSDP. We then used this framework to experiment with various distribution strategies and their impact on training performance. Starting with the results of these experiments, we also performed hyperparameter tuning to understand how these distribution strategies can work together to get high  GPU throughput on the Frontier system with AMD hardware and the ROCM software platform. With the lessons learned from these experiments, we further tuned the distribution strategies to develop recipes for training large models such as 175 billion and 1 trillion parameters. We achieved high GPU throughput and $100\%$ weak scaling efficiency for both models on thousands of GPUs. We also achieved $89\%$  and $87\%$ strong scaling efficiencies for these two models on thousands of GPUs.

One major challenge in reducing time-to-solution using a large part of Frontier will be loss divergence due to large batch size. To the best of our knowledge, the largest global batch size used in training an open-source LLM is 16 Million tokens, and most large LLMs have used much fewer tokens than the global batch size. With a sequence length of 2048, this translates to 8000 samples. We observed that at least one sample per GPU significantly boosts GPU throughput. To this extent, we must improve large-batch training and model parallel training performance with smaller per-replica batch sizes.

Most state-of-the-art distributed training frameworks target NVIDIA GPUs, and large-scale model training is done on CUDA-supported platforms. There needs to be more work exploring efficient training performance on AMD GPUs, and the ROCM platform is sparse. In this work, we have developed a training system of large LLMs of 175B and 1T on AMD hardware and the ROCM platform. This work can serve as the blueprint for efficient training of LLMs on non-NVIDIA and non-CUDA platforms such as AMD-powered Frontier supercomputer and Intel-powered Aurora supercomputer. 